\documentclass[nofootinbib,aps,pre,twocolumn,superscriptaddress]{revtex4}

\usepackage{amsmath,amssymb}
\usepackage{graphicx,color}
\usepackage{dcolumn}
\usepackage{mathrsfs}
\usepackage{bm}
\usepackage{bbm}
\usepackage{algorithm}
\usepackage{algpseudocode}
\usepackage{hyperref}
\algrenewcommand\algorithmicrequire{\textbf{Input:}}
\algrenewcommand\algorithmicensure{\textbf{Output:}}

\begin{document}

\title{Fickian yet non-Gaussian diffusion in an annealed heterogeneous
environment}

\author{Seongyu Park}
\affiliation{Department of Physics, Pohang University of Science and Technology
(POSTECH), Pohang 37673, Republic of Korea}
\author{Xavier Durang}
\affiliation{Department of Physics, Pohang University of Science and Technology
(POSTECH), Pohang 37673, Republic of Korea}
\author{Ralf Metzler}
\affiliation{Institute of Physics \& Astronomy, University of Potsdam, 14476
Potsdam-Golm, Germany}
\affiliation{Asia Pacific Center for Theoretical Physics (APCTP), Pohang 37673,
Republic of Korea}
\thanks{rmetzler@uni-potsdam.de}
\author{Jae-Hyung Jeon}
\thanks{jeonjh@postech.ac.kr}
\affiliation{Department of Physics, Pohang University of Science and Technology
(POSTECH), Pohang 37673, Republic of Korea}
\affiliation{Asia Pacific Center for Theoretical Physics (APCTP), Pohang 37673,
Republic of Korea}
\affiliation{School of Physics, Korea Institute for Advanced Study (KIAS), Seoul
130-722, Republic of Korea}

\date{\today}

\begin{abstract}
Fickian yet non-Gaussian diffusion is a ubiquitous phenomenon observed in
various biological and soft matter systems. This anomalous dynamics is
typically attributed to heterogeneous environments inducing spatiotemporal
variations in the diffusivity of tracer particles. While previous studies
have predominantly focused on systems exhibiting either spatial or temporal
heterogeneity, this work bridges the gap by introducing a model based on
an annealed extreme landscape to simultaneously account for both types
of heterogeneities. Through a combination of computational analyses and
analytical derivations, we investigate how the interplay of spatial and
temporal heterogeneities in the energy landscape gives rise to Fickian yet
non-Gaussian diffusion. Furthermore, we demonstrate that in the presence
of temporal environmental fluctuations, the heterogeneous diffusion
inevitably converges to classical Brownian motion via a homogenization
process. We derive an analytical expression for the homogenization time
as a function of key parameters governing the system's spatiotemporal
heterogeneities. Additionally, we quantify particle-to-particle diffusion
heterogeneity and examine the ergodic properties of this model, providing
deeper insights into the dynamics of complex, heterogeneous systems.
\end{abstract}

\maketitle

\section{Introduction}
\label{sec:intro}

Diffusion is a fundamental process that underpins various cellular and
biological phenomena. For instance, it plays a critical role in molecular
mechanisms such as the transcription and translation of DNA sequences
\cite{riggs1970,berg1981,park2021mini,pulkk}, cellular signaling
mediated by membrane proteins~\cite{jeon2016protein}, the transport
of mRNA--protein complexes~\cite{vargas2005mechanism,song2018neuronal,
lampo2017cytoplasmic}, as well as cellular motion and the transport of cargo
on cell carpets at longer times \cite{hartmutrev,beta,beta1}.
Recent advances in microscopy and single-particle
tracking techniques have enabled in-depth studies into the diffusion
dynamics of a wide range of biological tracers. These studies have
revealed that diffusion in biological systems often deviates from the
classical Brownian motion paradigm~\cite{hofling2013anomalous,metzler2014,pt,
munoz2021objective}. Notably, anomalous diffusion---characterized by a
power-law scaling of the mean squared displacement (MSD) with time of the form
$\langle x(t)^2\rangle\propto t^\alpha$ with $\alpha \neq 1$---has emerged as
a hallmark in these systems \cite{weiss2004anomalous,caspi2000enhanced,
metzler2014}. Subdiffusion with $0<\alpha<1$ is typically attributed to
environmental factors such as macromolecular crowding, viscoelastic properties
of the medium, or transient trapping induced by power-law, scale-free waiting
times \cite{goychuk2009viscoelastic,goychuk2012viscoelastic,scher1975anomalous}.

More recently, research has shifted focus toward a distinct phenomenon known
as Fickian yet non-Gaussian diffusion. This behavior, observed in diverse soft
and biological matter systems, is characterized by normal diffusion (linear
MSD scaling with time) but a substantial deviation from the Gaussian probability
density function (PDF) of the displacements \cite{wang2012brownian,
sabri2020elucidating,wang2009anomalous,ghosh2015non,xue2016probing,
cherstvy2019non,chakraborty2020disorder}. Instead, these systems exhibit
non-Gaussian PDFs, often resembling Laplace distributions of the form
\begin{eqnarray}
p(x,t)\propto\exp\left(-|x|/\lambda(t)\right),
\end{eqnarray}
with the time-dependent width $\lambda(t)$ that may scale as $\lambda(t)
\simeq t^{1/2}$ or $\simeq t^{1/3}$ \cite{chechkin2017,roldan}. Additional
power-law corrections may modify the Laplace shape, especially around the
cusp \cite{chechkin2017,sposini2018,wang2012brownian}. The origin of this
non-Gaussianity is typically attributed to environmental heterogeneities
\cite{sabri2020elucidating,chechkin2017,chubynsky2014diffusing,lanoiselee2018,
jeon2016protein,sposini2018,cherstvy2019non,thapa2018bayesian,
lanoiselee2018model} or the conformational variability of the tracer particles
\cite{uneyama2015fluctuation,yamamoto2021universal,miyaguchi2017elucidating,
kamagata2018high}. In many cases, fluctuations in particle diffusivity are
described by a static distribution of diffusivity values, $\psi(D)$, leading to
the superstatistical formulation for the PDF \cite{beck2003superstatistics}
\begin{eqnarray}
\label{eq:superstatistics_intro}
p(x,t)=\int_0^{\infty}\mathrm{d}D~\psi(D)G(x,t|D),
\end{eqnarray}
where $G(x,t|D)=(4\pi Dt)^{-1/2}\exp(-x^2/[4Dt])$ is a Gaussian propagator for
a given diffusivity $D$. Certain formulations of diffusing-diffusivity models
have the superstatistical formulation as their short-time limit, see, e.g.,
\cite{chechkin2017,sposini2018}.

The description of fluctuating diffusivities varies with the temporal and
spatial characteristics of the system. For environments with rapid temporal
fluctuations relative to the particle dynamics, the diffusivity can be treated
as a stochastic, exclusively time-dependent variable $D(t)$
\cite{chubynsky2014diffusing,chechkin2017,thapa2022bayesian,lanoiselee2018,
lanoiselee2018model,sposini2018}, which in the area of stochastic processes
is often referred to as \emph{annealed disorder\/} \cite{bouchaud,stas}. A
prominent example of the
annealed disorder is the diffusing diffusivity model suggested by Chubynsky
and Slater~\cite{chubynsky2014diffusing}, which was further elaborated by
Chechkin \emph{et al}~\cite{chechkin2017}. Conversely, for systems with
slow environmental changes, diffusivity is better represented as a spatially
dependent variable $D(\mathbf{r})$~\cite{postnikov2020,luo2018,luo2019},
which is called \emph{quenched disorder\/} \cite{bouchaud,stas}. In quenched
systems, typically, the diffusivity has the same value each time the particle
(re)visits the same location in space, leading to the buildup of correlations.

However, many biological systems exhibit more complex spatiotemporal
heterogeneities, where neither annealed nor quenched disorder alone
suffices to adequately describe the observed diffusive dynamics. Examples
include crowded cellular membranes \cite{jeon2016protein,metzler2016non,
weigel2011ergodic,he_nc}, dynamic chromatin
structures~\cite{di2018anomalous, sung2021stochastic, saintillan2018extensile,
mahajan2022euchromatin, zidovska2013micron, banigan2020loop, kim2019human,
hansen2018recent, joo2020anomalous}, crowded polydisperse vacuole
systems~\cite{thapa2019transient, reverey2015superdiffusion}, and
active cytoskeletal networks~\cite{guo2014probing, joo2020anomalous,
sungkaworn2017single, sadegh2017plasma, moore2016dynamic}.

In this study, we explore the Fickian yet non-Gaussian diffusion of
tracer particles in a novel spatiotemporally heterogeneous medium, the
\emph{annealed extreme landscape}. This framework integrates both spatial
and temporal dependencies of diffusivity, capturing the intricate interplay
between environmental heterogeneity and particle dynamics. By combining
analytic derivations and computational simulations, we investigate key
diffusion metrics, including MSDs, Van-Hove self-correlation functions,
non-Gaussianity, and ergodicity breaking parameters. Our analysis elucidates
how these quantities evolve with varying environmental fluctuation scales and
reveals the homogenization process as a consequence of which non-Gaussian
diffusion eventually cross over to Gaussian behavior. Through these
investigations, we aim to provide a comprehensive understanding of diffusion
dynamics in complex, heterogeneous media.

This work is organized as follows. Section~\ref{sec2} introduces the
annealed extreme landscape model, outlining several dynamic observables used
to quantify the spatiotemporal heterogeneity of the energy landscape. In
Sec.~\ref{sec:nongaussian}, we investigate the diffusion dynamics in the
annealed extreme landscape, focusing on quantities such as MSD, the Van-Hove
self-correlation function, and measures of non-Gaussianity. Furthermore,
we observe the Fickian yet non-Gaussian diffusion and provide an analytical
framework to explain thus effect under quenched and fast-annealing conditions.
Section~\ref{sec:homogenization} delves into the homogenization process,
deriving the analytic expression for the homogenization time based on the
self-similarity of the random energy landscape. Additionally, we explore
particle-to-particle diffusion heterogeneity via the ergodicity-breaking
parameter. Finally, Sec.~\ref{sec:discussion} concludes by summarizing
the key findings and discussing their implications for the understanding
of Fickian yet non-Gaussian diffusion phenomena.

\section{Annealed extreme landscape model}
\label{sec2}

In this Section, we model the spatiotemporal heterogeneous diffusion of
tracers embedded in an environment whose randomness changes over time. We
describe such random media using a two-dimensional lattice model called the
annealed extreme landscape. Here, we encapsulate the essence of this model
along with the introduction of associated dynamic observables we developed,
which are call the local diffusivity field (Sec.~\ref{sec:2B}) and the
sampled diffusivity (Sec.~\ref{sec:2C}).

\subsection{Annealed extreme landscape}
\label{sec:annealed_field}

We extend the quenched extreme landscape model introduced in
Refs.~\cite{luo2018,luo2019} to incorporate temporal fluctuations. The
annealed extreme landscape is constructed by the following steps:

\begin{figure*}
\centering
\includegraphics[width=0.85\textwidth]{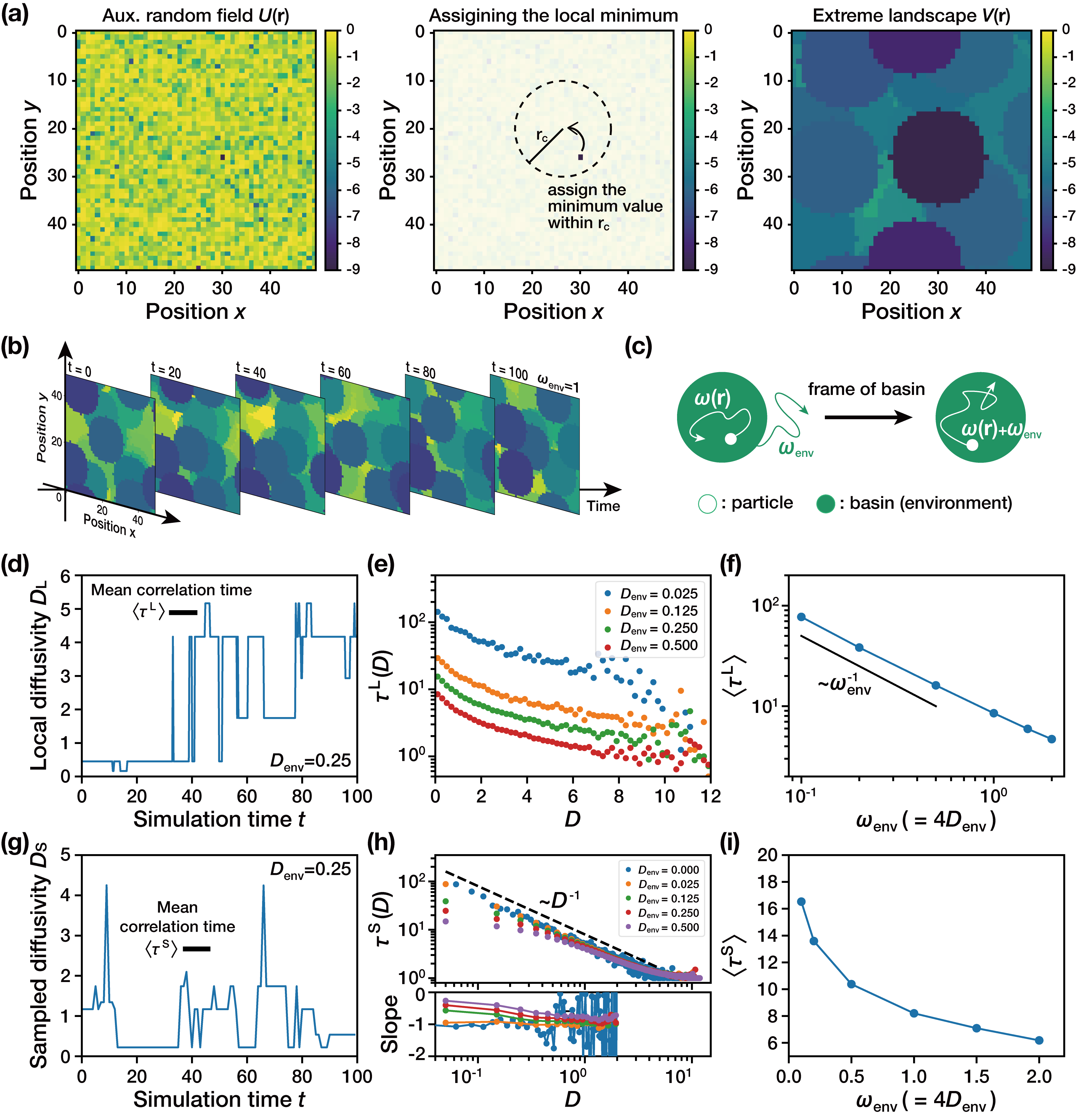} 
\caption[An annealed extreme landsape and its dynamic properties]{\textbf{An annealed extreme landscape and its dynamic properties.}
(a) Generation of an extreme landscape. In the left panel, an auxiliary random field is generated (with $L=50$ in this figure) by assigning random energies to the lattice points, sampled from the exponential distribution~\eqref{eq:auxiliary}.
In the middle panel, for a given auxiliary field, the local energy minimum within a critical radius $r_\mathrm{c}$ is assigned to every lattice point. Finally, in the right panel, the resulting extreme landscape is constructed. (b) Time evolution of the
annealed extreme landscape shown in (a). Here, $\omega_\mathrm{env}=1$. (c)
Schematic illustrating a tracer particle diffusing in a basin of an annealed
extreme landscape. The particle moves to an adjacent lattice point with the
rate $\omega(\mathbf{r})$, while the environment changes with the rate $\omega
_{\mathrm{env}}$. In the reference frame of the basin (right), a particle
diffuses with the combined rate $\omega(\mathbf{r})+\omega_{\mathrm{env}}$
while the environment remains immobile. (d) An example of the time evolution
of the local diffusivity at a fixed position $\mathbf{r}$ with $D_\mathrm{env}
=0.25$. The bold black line represents the duration of the mean residence time
$\langle\tau^\mathrm{L}\rangle=8.52$. (e) PDF of residence times $\tau^\mathrm{
L}$ of the local diffusivity $D$ for various values of $D_\mathrm{env}$. (f)
Mean residence times of the local diffusivities, $\langle\tau^\mathrm{L}\rangle$,
as a function of $\omega_\mathrm{env}$. The solid line depicts the scaling with
the power-law $1/\omega_\mathrm{env}$. (g) A simulated sampled diffusivity $D_S$
for $D_\mathrm{env}=0.25$. The bold black line represents the mean correlation
time $\langle\tau^\mathrm{S}\rangle=8.20$. (h) The distribution of residence
times $\tau^\mathrm{S}$ of the sampled diffusivities as a function of $D$ for
several values of $D_\mathrm{env}$. The dashed line denotes the power-law scaling
$1/D$. The bottom panel shows the fitted power-law slope for $\tau^\mathrm{S}(D)$.
(i) Mean residence times $\langle\tau^\mathrm{S}\rangle$ of the sampled
diffusivities, as a function of $\omega_\mathrm{env}$.}
\label{fig1}
\end{figure*}

(i) \textit{Auxiliary random field.} First, consider a two-dimensional
square lattice of size $L\times L$ with the lattice constant $a=1$. Every
lattice point $\mathbf{r}=(x,y)$ is allocated the auxiliary random energy
$U(\mathbf{r})$, drawn from the exponential PDF [Fig.~\ref{fig1}(a)]
\begin{eqnarray}
\label{eq:auxiliary}
\phi_U[U(\mathbf{r})=U]=U_0^{-1}\exp\left(\frac{U}{U_0}\right),\quad U<0.
\end{eqnarray}
We refer to the set of random variables $\{U(\mathbf{r})\}$ as an auxiliary
random field. In our study, we set the parameter $U_0$ to unity for simplicity
and typically consider a random field of size $L=512$, unless specified
otherwise.

(ii) \textit{Extreme landscape.} Once the auxiliary random field
$U(\mathbf{r})$ is given, the extreme landscape $V(\mathbf{r})$ is defined
at each lattice point $\mathbf{r}$ as the local minimum value among the
neighboring values $U(\mathbf{r}')$ within a critical radius $r_c$ (as shown in the middle panel of Fig.~\ref{fig1}(a)). This is
mathematically expressed as
\begin{eqnarray}
\label{eq:extreme_landscape}
V(\mathbf{r})=\min[U(\mathbf{r}')\,|\,\|\mathbf{r}-\mathbf{r}'\|\leq r_c].
\end{eqnarray}
In a finite lattice space, periodic boundary conditions are employed to
construct $V(\mathbf{r})$. When $r_c$ is sufficiently large, the distribution
of $V(\mathbf{r})$ is known to obey the Gumbel distribution
\cite{luo2018,luo2019}
\begin{eqnarray}
\label{eq:gumbel}
\phi_V[V(\mathbf{r})=V]\approx\exp\Big[V-V_0-\exp(V-V_0)\Big]
\end{eqnarray}
where $V<0$ and $V_0=-\log(\pi r_c^2/a^2)$. Throughout our work, we set the
critical radius to be $r_c=10$ in our simulations. The right panel of Fig.~\ref{fig1}(a)
illustrates an example of the computer-generated extreme landscape.

(iii) \textit{Annealing.} For a given time interval $\delta$, we randomly
choose $L^2\omega_\mathrm{env}\delta/2$ lattice points (where $\omega_\mathrm{
env}$ is the annealing rate per unit time $t_0(=1)$) and update their auxiliary
random fields according to the following rule: For a chosen lattice point
$\mathbf{r}$ and one of its four nearest neighbors $\mathbf{r}'$, we exchange
their respective random fields $U(\mathbf{r})$ and $U(\mathbf{r}')$ with each
other. Note that $\omega_\mathrm{env}$ can be interpreted as the average number
of exchange events per lattice point per unit time. For a more detailed
description of the algorithm, we refer to App.~\ref{sec:appendixA}.

(iv) Repeat the above processes (ii) and (iii) every time interval $\delta$
to realize the annealed extreme landscape $V(\mathbf{r},t)$, that then
becomes a fluctuating field over time $t$.

As a result of the above workflow, we obtain the annealed extreme landscape
with the environmental diffusivity
\begin{eqnarray}
\label{eq:D_env}
D_\mathrm{env}\equiv\frac{\omega_\mathrm{env}a^2}{4t_0}=\frac{\omega_\mathrm{
env}}{4}.
\end{eqnarray}
Figure~\ref{fig1}(b) depicts an exemplary annealed extreme landscape that
evolves over time (here, $\omega_\mathrm{env} = 1$ and $D_\mathrm{env} = 0.25$).

\subsection{Local diffusivity field}
\label{sec:2B}

To describe the local trapping/escape dynamics on a given annealed extreme
landscape $V(\mathbf{r},t)$, we trace the diffusing dynamics of a tracer
particle. The Kramers escape rate from the local trapping site $\mathbf{r}$ at
time $t$ is given by
\begin{eqnarray}
\label{eq:walkrate}
\omega(\mathbf{r},t)=\omega_0\exp\left[V(\mathbf{r},t)\right],
\end{eqnarray}
where $\omega_0=4\pi r_c^2/(a^2t_0)$ and, for simplicity, the ambient temperature $k_\mathrm{B}\mathrm{T}$ is set to
unity. Our choice of $\omega_0$ sets the escape rate at a local trap of depth
$V_0$ to 4 per unit time $t_0(=1)$, and the corresponding local diffusivity of the tracer becomes unity [$a^2/t_0$].
We find that the Kramers escape rates from the trapping sites follow the
exponential law (see the derivation in App.~\ref{sec:appendixB})
\begin{eqnarray}
f_\omega[\omega(\mathbf{r},t)=\omega]=\frac{t_0}{4}\exp\left( -\frac{\omega
t_0}{4}\right).
\end{eqnarray}
After the escape, the particle hops onto one of its four nearest neighbor sites
with equal probability of $1/4$.

In Fig.~\ref{fig1}(c), we schematically illustrate the diffusion of a tracer
particle (white dot) within an annealed extreme landscape. The particle hops
onto the lattice with a jump rate of $\omega(\mathbf{r}, t)$ while the extreme
landscape changes over time with the annealing rate $\omega_\mathrm{env}$
(or the annealing diffusivity $D_\mathrm{env}$) [Eq.~\eqref{eq:D_env}].
As these two contributions are independent of each other, the tracer's
positional net change relative to an extreme basin occurs at a rate of
$\omega(\mathbf{r},t)+\omega_\mathrm{env}$. Alternatively, the sojourn time
of a particle at site $\mathbf{r}$ within an extreme basin is given by
\begin{eqnarray}
\label{eq:tau}
\tau(\mathbf{r},t)=\frac{1}{\omega(\mathbf{r},t)+\omega_\mathrm{env}}.
\end{eqnarray}
Using a change of variables (App.~\ref{sec:appendixB}), the PDF of sojourn times
$\tau\in(0,\omega_\mathrm{env}^{-1})$ is obtained as
\begin{eqnarray}
f_\tau[\tau(\mathbf{r},t)=\tau]=\frac{\tau_0 \exp(\omega_\mathrm{env}\tau_0)}
{\tau^2}\exp\left(-\frac{\tau_0}{\tau}\right),
\end{eqnarray}
where $\tau_0=t_0/4$. Accordingly, the stationary positional PDF of tracer
particles is given by
\begin{eqnarray}
\label{eq:stationary}
\pi(\mathbf{r},t)=\frac{\tau(\mathbf{r},t)}{\sum_{\mathbf{r}'}\tau(\mathbf{r}',
t)}
\end{eqnarray}
in terms of the sojourn time PDF \eqref{eq:tau}.

We then define the local diffusivity of the tracer particles in the annealed
extreme landscape as
\begin{eqnarray}
\label{eq:localdiff}
D_\mathrm{L}(\mathbf{r},t)=\frac{\omega(\mathbf{r},t)a^2}{4}=\pi r_c^2\exp
\left[V(\mathbf{r},t)\right].
\end{eqnarray}
This relation indicates that the local diffusivity field is solely determined
by the spatiotemporal heterogeneity of the extreme landscape. The PDF of local
diffusivities on the annealed extreme landscape reads
\begin{equation}\label{eq:localdiff_exact}
\begin{aligned}
\psi_\mathrm{L}(D_\mathrm{L}=D)=\frac{1}{D_0}\exp\left(-\frac{D}{D_0}\right),
\end{aligned}
\end{equation}
as obtained via Eq.~\eqref{eq:gumbel}, where $D_0=\omega_0a^2\exp(V_0)/4=1$.

In Fig.~\ref{fig1}(d), we plot the time evolution of the local diffusivity
$D_\mathrm{L}(\mathbf{r},t)$ at a fixed point $\mathbf{r}$ from stochastic
simulations. It is observed that the local diffusivity exhibits significant
fluctuations in time due to the annealed dynamics of the extreme landscape.
Using the time traces of the simulated local diffusivities, we compute the
average residence time for a given value of $D_\mathrm{L}$. Figure~\ref{fig1}(e)
presents the relation of the residence time vs.~$D_\mathrm{L}$. We find that a
larger value $D_\mathrm{L}$ (associated with a shallower $V$) has a shorter
residence time $\tau^\mathrm{L}$. This inverse behavior stems from the fact
that the shallow potential traps, which are overshadowed by the deeper ones,
have smaller spatial correlation lengths, as illustrated in Fig.~\ref{fig1}(a)
or (b), and the particle diffuses faster over these shallow potential regions.
In Fig.~\ref{fig1}(f), we plot the mean residence time $\langle\tau^\mathrm{L}
\rangle$ against the annealing rate. The mean residence time $\langle \tau^
\mathrm{L}\rangle$ is found to be, approximately, inversely proportional to
$\omega_\mathrm{env}$.

\subsection{Sampled diffusivity}
\label{sec:2C}

We now define the sampled diffusivity $D_\mathrm{S}(t)$ as the time sequence
of the local diffusivities that a diffusing tracer visits over time; see
Fig.~\ref{fig1}(g) for an example. As the tracer particle diffuses on a
random landscape, $D_\mathrm{S}(t)$ in general is temporally fluctuating and
reflecting the correlated nature of the local diffusivity field. In this sense,
the stochastic behavior of the sample diffusivity is reminiscent of the
diffusivity dynamics in the fluctuating diffusivity model \cite{sposini2018,
chechkin2017,postnikov2020} or the annealed transit time model (ATTM)
\cite{massignan2014nonergodic}.

In Fig.~\ref{fig1}(h), we numerically estimate the relationship between the
residence time $\tau^S(D)$ of the sampled diffusivity against its value
$D$. For various annealing rates, a general tendency is that the larger
the sampled diffusivity $D$ the shorter its residence time $\tau^S(D)$.
In particular, in the limit of a fully quenched landscape ($D_\mathrm{env}\to
0$), the residence time PDF has the power-law scaling $\tau^\mathrm{S}(D)
\sim D^{-\gamma}$ with $\gamma\approx1$ within our observation time window. A
similar relation between the diffusivity and its residence time is defined in
the annealed transit time model (ATTM) in which the residence time for a given
diffusivity $D$ follows a power-law relation with the expectation $\mathbb{E}
[\tau|D]\sim\tau^{-\gamma}$~\cite{massignan2014nonergodic}. As the annealing of the system gets more
pronounced ($\omega_\mathrm{env}\gg1$), the overall behavior of $\tau^\mathrm{
S}$ deviates from the simple power-law scaling, because the annealing induces
the release of a particle from a deep local trap.

\begin{figure}
\includegraphics[width=0.38\textwidth]{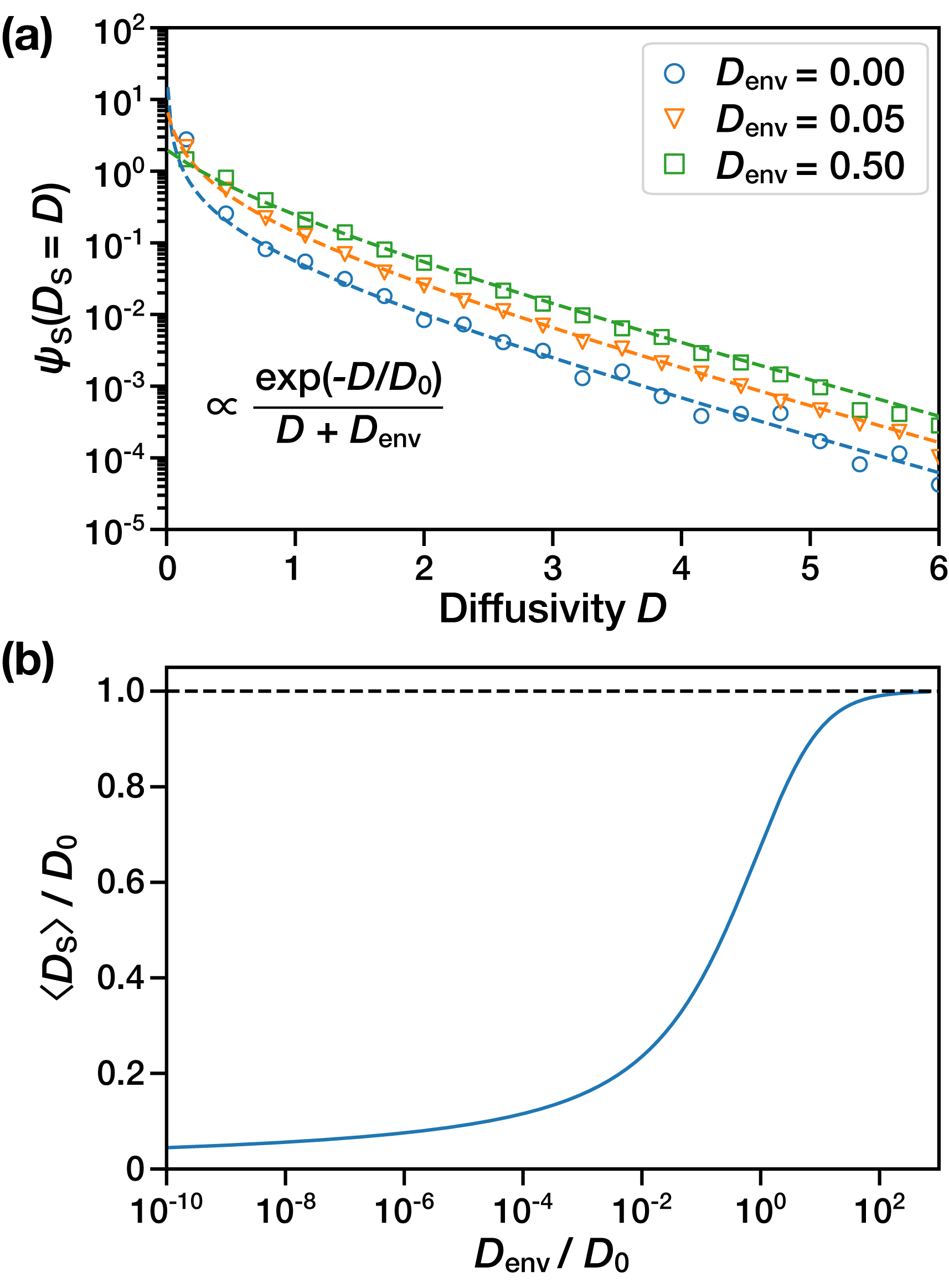} 
\caption{\textbf{Profiles and the mean values of the sampled diffusivities.}
(a) Normalized PDF $\psi_\mathrm{S}(D)$ of the sampled diffusivities for
various values of $D_{\mathrm{env}}$. The simulations results (symbols) are
in excellent agreement with the theoretical PDF \eqref{eq:sampled_D_dist}
depicted as the dashed lines. (b) Theoretical result
[Eq.~\eqref{eq:mean_sampled_diffusivity}] for the mean sampled diffusivities
$\langle D_\mathrm{S}\rangle/D_0$ as function of $D_\mathrm{env}$. The dashed
line shows the limiting value for $D_\mathrm{env}\rightarrow\infty$.}
\label{fig2}
\end{figure}

The relationship between the PDFs $\psi_\mathrm{S}(D_\mathrm{S})$ of the
sampled diffusivity and $\psi_\mathrm{L}(D_\mathrm{L})$ of the local
diffusivity can be derived as follows. Given that a particle sits on a site
with $D_\mathrm{L}=\frac{\omega a^2}{4}$, the particle's sojourn time at this
site is $\tau=1/(\omega+\omega_\mathrm{env})$ [Eq.~\eqref{eq:tau}]. Therefore,
the PDF of the sampled diffusivity $D_\mathrm{S}=D$ is obtained as
\begin{equation}
\label{eq:sampled_D_dist}
\begin{aligned}
\psi_\mathrm{S}(D_\mathrm{S}=D)&\propto\frac{1}{\omega+\omega_\mathrm{env}}
\times\text{total area associated with }D\\
&\propto\frac{\psi_\mathrm{L}(D_\mathrm{L}=D)}{D+D_\mathrm{env}}.
\end{aligned}
\end{equation}
We note that in the quenched limit ($D_\mathrm{env}\to0$), the above
relationship recovers the result reported in Ref.~\cite{postnikov2020} for
the quenched extreme landscape,
\begin{eqnarray}
\label{eq:quenched}
\psi_\mathrm{S}(D_\mathrm{S}=D)\propto\frac{\psi_\mathrm{L}(D)}{D}.
\end{eqnarray}
In Fig.~\ref{fig2}(a), we show the distribution of the sampled diffusivity
acquired from stochastic simulations (symbols) for several values of
$\omega_\mathrm{env}(=4D_\mathrm{env}/a^2)$. The simulated data is in
excellent agreement with the expected sampled diffusivity PDF
\eqref{eq:sampled_D_dist}, depicted as the dashed lines.

A notable effect of the environmental change (quantified by $D_\mathrm{env}$)
is the reduced probability for sampling small-valued diffusivities. In the
quenched landscape, a particle spends a long time escaping from a region
of slow diffusivity. In contrast, in an annealed environment, a trapped
particle can escape more quickly from such regions because the environment
itself changes. Consequently, the mean sampled diffusivity of a particle
increases in an annealed landscape. The mean sampled diffusivity $\langle
D_\mathrm{S}\rangle$ of a particle is obtained as
\begin{equation}
\label{eq:mean_sampled_diffusivity}
\langle D_\mathrm{S}\rangle=D_0\frac{\exp\left(-D_\mathrm{env}/D_0\right)}{
\Gamma\left(0,D_\mathrm{env}/D_0\right)}-D_\mathrm{env},
\end{equation}
where $\Gamma(s,x)=\int_x^\infty t^{s-1}e^{-t}dt$ is the upper incomplete Gamma function~\cite{abramowitz1964handbook}.
In Fig.~\ref{fig2}(b), we plot the mean sampled diffusivity $\langle D_\mathrm{
S}\rangle/D_0$ against $D_\mathrm{env}/ D_0$. Notably, the mean diffusivity
$\langle D_\mathrm{S}\rangle$ monotonically increases with increasing $D_\mathrm{
env}$, converging to $D_0$ in the large-$D_\mathrm{env}$ limit. This trend is
indeed observed in our simulation video \cite{sm}, where the tracers attain a
faster diffusivity as $D_\mathrm{env}$ increases. We also note that the mean
sampled diffusivity converges to zero in the quenched limit $D_\mathrm{env}\to0$
[Eq.~\eqref{eq:mean_sampled_diffusivity}]. This happens because the diffusing
tracers end up highly immobilized in regions where $V(\mathbf{r})\to-\infty$.
However, such an infinite trap cannot be realized in a finite system ($L<
\infty$), in which the particles remain mobile even when $D_\mathrm{env}$ is
negligible (but finite). In previous studies on the quenched extreme landscape
\cite{luo2018, luo2019}, it was demonstrated that the longest sojourn time of a
diffusing tracer scales as $\tau\sim L^2$ and the mean sampled diffusivity is
inversely proportional to the logarithm of the system size, $\langle D_\mathrm{S}
\rangle\propto 1/\log L^2$.

\section{Fickian yet non-Gaussian diffusion}
\label{sec:nongaussian}

\subsection{Mean squared displacements}

To characterize the diffusion dynamics of tracers in extreme energy landscapes,
we first estimate the MSDs from the simulated trajectories. Figs.~\ref{fig8}(a),
(b) show the MSDs of tracers in quenched and annealed landscapes, respectively.
Each case was simulated with stationary initial positions of tracers according
to Eq.~\eqref{eq:stationary} for $N_\mathrm{par}=1000$ tracers. In the plot,
the gray solid lines represent 500 time-averaged (TA) MSD curves from individual
trajectories according to the definition \cite{pt1,metzler2014}
\begin{eqnarray}\label{eq:tamsd}
\overline{\delta^2(\mathit\Delta)}=\frac{1}{T-\mathit\Delta}\int_0^{T-\mathit
\Delta}\mathrm{d}t\,\left[x(t+\mathit\Delta)-x(t)\right]^2,
\end{eqnarray}
and the thick line (black) represents their ensemble-average (EA), $\left<
\overline{\delta^2(\mathit\Delta)}\right>$, the EATAMSD. Here, $\mathit\Delta$
represents the lag time, the magnitude of the sliding window, and $T$ stands
for the total observation time or length of the time series $x(t)$. For
comparison, in Fig.~\ref{fig8}, we also calculate the conventional (EA)
MSD\footnote{In the following referred to simply as the MSD.}
\begin{eqnarray}
\label{eq:eamsd}
\langle\Delta x(t)^2\rangle=\frac{1}{N_\mathrm{par}}\sum_{i=1}^{N_\mathrm{par}}
[x_i(t)-x_i(0)]^2
\end{eqnarray}
shown by the circles ($\circ$). 

In both the quenched ($D_\mathrm{env}=0$) and annealed ($D_\mathrm{env}=0.5$)
cases, the EATAMSD and MSDs coincide with each other and increase linearly
with time throughout the observation time window. These results suggest that
the tracers in heterogeneous random environments exhibit Fickian diffusion
regardless of the detail of the spatiotemporal heterogeneity. We note that
the individual TAMSDs also follow the linear time evolution and that their
amplitude scatter is very narrow (apart from long lag times $\Delta\to T$ when
the statistic naturally deteriorates). This latter behavior is obtained when
the total observation time $T$ is sufficiently longer than the so-called
homogenization time $\tau_c$, that will be introduced below in
Sec.~\ref{sec:homogenization}. In the opposite regime $T\ll\tau_c$, the
tracers explore locally distinct random environments, leading to noticeable
trajectory-to-trajectory amplitude scatter in the TAMSD. For instance, see
the MSDs for $T=2\times10^3$ ($<\tau_c\approx10^4$) in Fig.~\ref{fig8}a
(which will be discussed in detail below).

The observed Fickian diffusion can be explained as follows: At short times
$t'\ll\langle\tau^\mathrm{S}\rangle$, the diffusivity $D_\mathrm{S}$ sampled
by a particle remains approximately a constant, such that the diffusion is
Fickian and $\mathbb{E}[[x(t')-x(0)]^2]\approx2D_\mathrm{S}t'$. Therefore, the
MSD from a collection of tracers for the time range $t'$ can be simply
written as
\begin{eqnarray}
\nonumber
\langle[x(t')-x(0)]^2\rangle&\approx&2\langle D_\mathrm{S}\rangle t'\\
&=&2a^2
\left<\frac{\omega t'}{4}\right>=\frac{a^2}{2}\langle n(t')\rangle.
\label{eq:eamsd_vs_steps}
\end{eqnarray}
Here, $n(t')$ denotes the number of steps of the random walk performed by
the particle during the time interval $[0,t']$, and the average $\langle
\cdots\rangle$ is taken over an ensemble of the stationary initial condition.
Given that the ensemble is in the stationary state, the MSD over the time
interval $[t',2t']$ is also given by $\langle[x(2t')-x(t')]^2\rangle\approx
a^2\langle n(t')\rangle/2$, meaning that the same number of random walk
steps approximately occur during $[t',2t']$. Therefore, for an arbitrary $t>
0$, the number of random walk steps during the time interval $[0,t]$ is
$\langle n(t)\rangle=\langle n(t')\rangle\times(t/t')$. Plugging this
relation into Eq.~\eqref{eq:eamsd_vs_steps}, the MSD at time $t$ is obtained
as
\begin{eqnarray}
\label{eq:fickianlaw}
\langle\Delta x(t)^2\rangle\approx2\langle D_\mathrm{S}\rangle t.
\end{eqnarray}
This result suggests that for stationary initial conditions the MSD (along
with the EATAMSD) satisfy the Fickian diffusion law with the diffusivity
given by the ensemble-averaged sampled diffusivity $\langle D_\mathrm{S}
\rangle$. This is indeed confirmed in Fig.~\ref{fig8}(b) where our theory
Eq.~\eqref{eq:fickianlaw} perfectly explains the simulated MSD and EATAMSD
in the annealed random system ($D_\mathrm{env}\neq0$). In the case of
a non-stationary initial condition, the MSDs exhibit short-time Fickian
diffusion with a diffusivity that depends on the initial distribution
of the sampled diffusivity. Thus, in general, the non-stationary Fickian
diffusion observed is different from Eq.~\eqref{eq:fickianlaw}. Moreover,
a non-Fickian diffusion emerges during the cross-over time before converging
to the stationary behavior \eqref{eq:fickianlaw} after homogenization.
Further discussion on this point is presented in App.~\ref{sec:nonstationary},
see Fig.~\ref{fig:nonstationary}.

\begin{figure}
\centering
\includegraphics[width=0.38\textwidth]{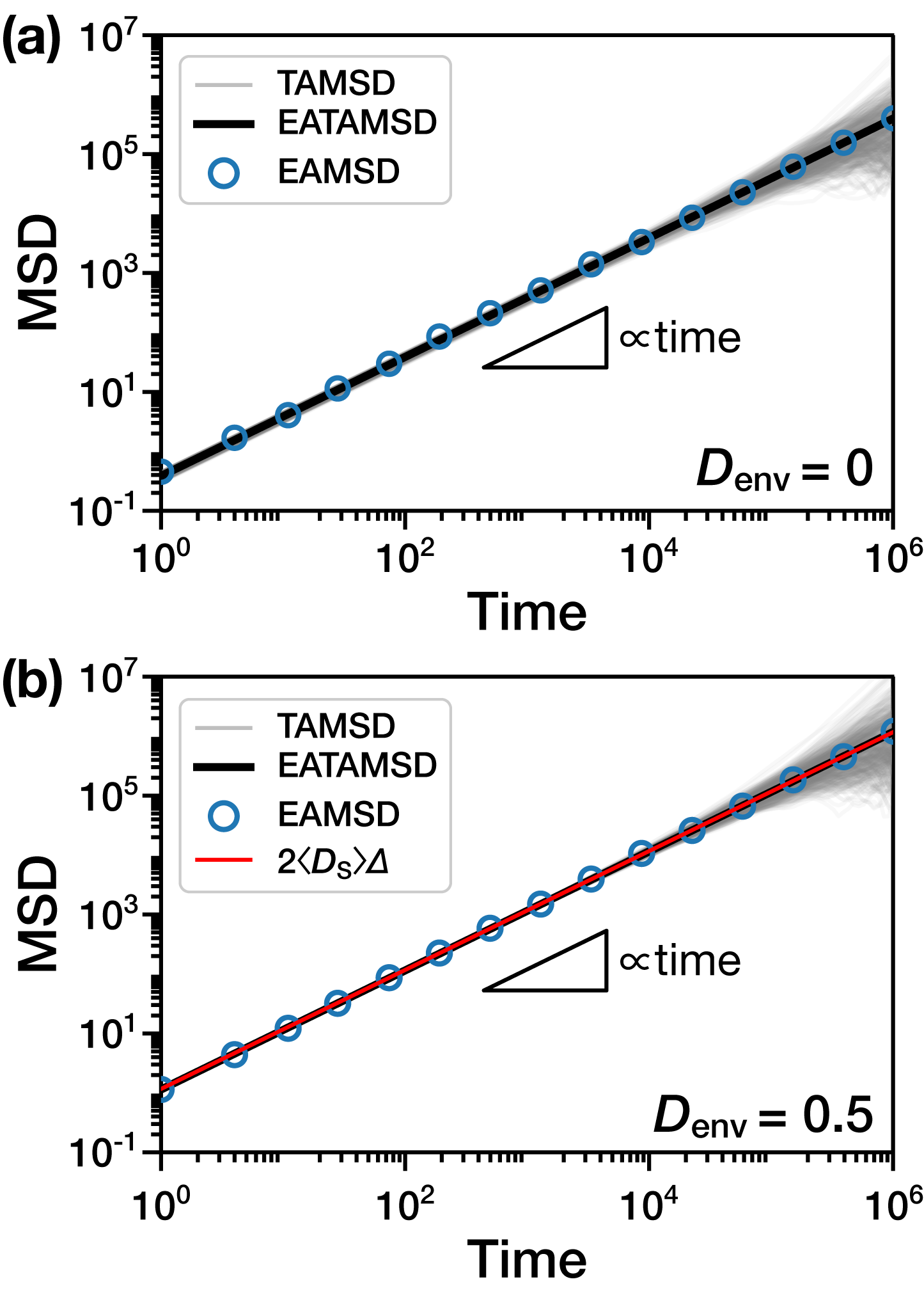} 
\caption{\textbf{Mean squared displacement estimates from stochastic simulations.}
(a) MSDs from simulated trajectories ($N_\mathrm{par}=1000$, $T=2\times10^6$) for
the quenched extreme landscape ($D_\mathrm{env}=0$). Individual TAMSDs (grey
lines), MSD (blue circles), and EATAMSD (black line) overlap nicely and exhibit
Fickian (normal) diffusion. (b) MSDs ($N_\mathrm{par}=1000$, $T=2\times10^6$) for
an annealed extreme landscape ($D_\mathrm{env}=0.5$). The individual TAMSDs
(grey lines), MSD (blue circles), and EATAMSD (black line) again exhibit
Fickian behavior, as predicted by the theoretical result \eqref{eq:fickianlaw}.
For the simulations in the panels of (a) and (b), the initial distribution of
the tracer particles was chosen according to the stationary distribution given
by Eq.~\eqref{eq:stationary}.}
\label{fig3}
\end{figure}

\subsection{Van-Hove self-correlation function}
\label{sec:3b}

We next examine the distribution of tracer displacements in both the quenched
and annealed landscapes studied above. For each case, the Van-Hove
self-correlation function $p(x,\mathit{\Delta})$ is calculated for the ensemble
of tracers at a short ($\mathit{\Delta}=1$) and long ($\mathit{\Delta}=10^4$)
lag times and shown in Fig.~\ref{fig3}. We note that while the MSDs are Fickian,
the heterogeneous diffusion of the tracers is highly non-Gaussian. In the
quenched environment, the Van-Hove self-correlation function has a sharp
cusp at the center and decays exponentially. The cusp originates from immobile
tracers trapped in deep potential wells. The escape of these tracers from the
traps occurs at long time scales. Thus, the cusp remains persistent even when
the lag time becomes substantial ($\mathit{\Delta}=10^4$).

\begin{figure}
\includegraphics[width=0.48\textwidth]{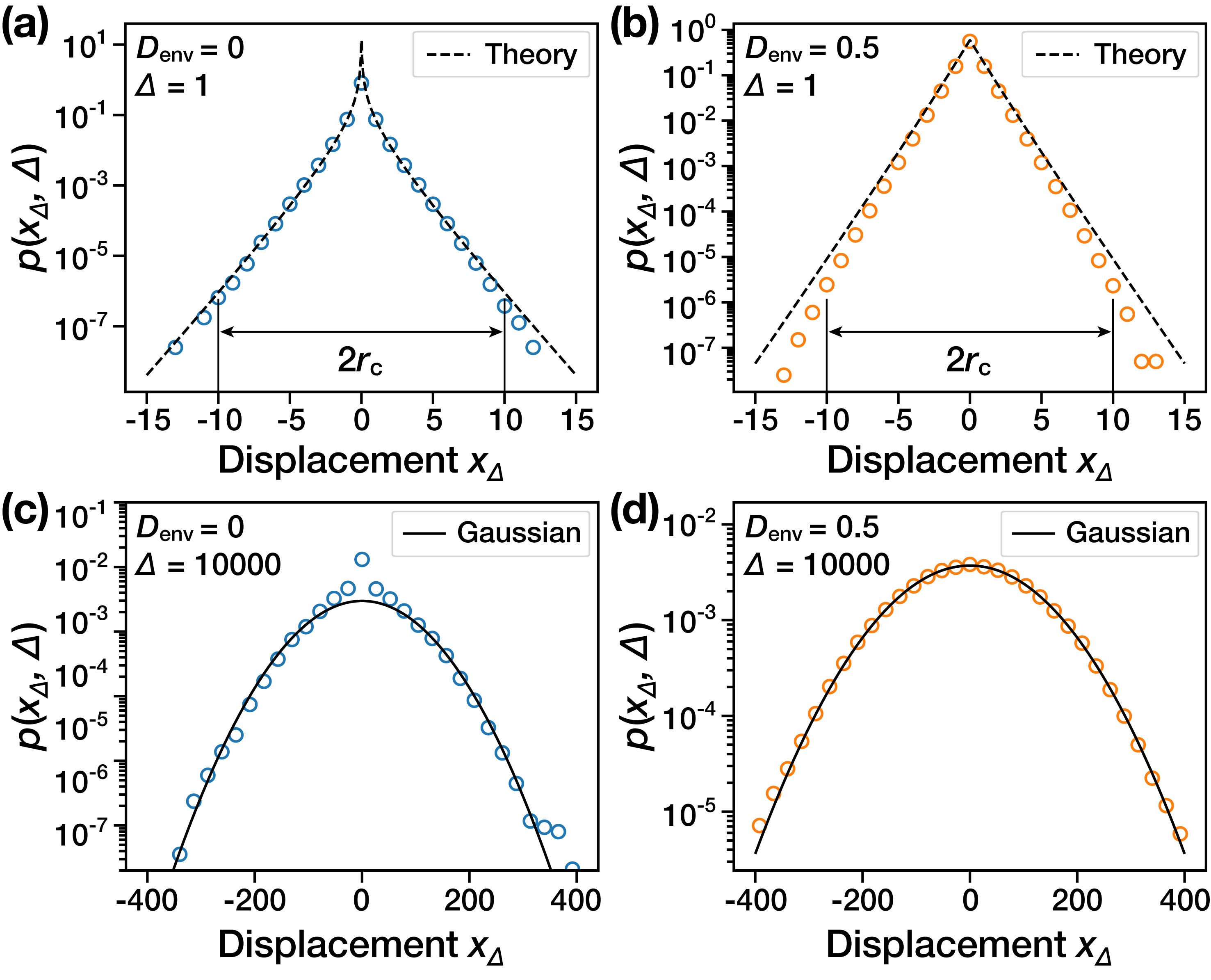} 
\caption{\textbf{Van-Hove self-correlation functions from the simulated
trajectories.} (a) and (b): Van-Hove self-correlation function extracted
from stochastic simulations (symbols, $N_\mathrm{par}=1000$) at lag time
$\mathit\Delta=1$ for (a) a quenched extreme landscape ($D_\mathrm{env}=0$)
and (b) an annealed extreme landscape ($D_\mathrm{env}=0.5$). The black
dashed lines denote the analytic result \eqref{eq:exact}. (c) and (d):
Van-Hove self-correlation function (symbols, $N_\mathrm{par}=1000$) at lag
time $\mathit\Delta=10^4$ for (c) a quenched extreme landscape ($D_\mathrm{
env}=0$) and (d) an annealed extreme landscape ($D_\mathrm{env}=0.5$). In
both panels (c) and (d), the black solid lines show the best Gaussian fit to
the data.}
\label{fig4}
\end{figure}

Notably, in the annealed random environment, the tracer diffusion exhibits
a non-Gaussian behavior analogous to the fluctuating diffusivity model
\cite{chechkin2017}. On short time scales, the tracer displacement follows
the Laplace distribution (shown here for $\mathit{\Delta}=1$). The Laplace
Van-Hove self-correlation function crosses over to a Gaussian at long time
scales when $t\gtrsim\tau_c$. In Fig.~\ref{fig3}](d), the simulation data is
perfectly fitted by a Gaussian. Summarizing the results of the MSD and
displacement PDF, we find that in an annealed extreme landscape the tracer
exhibits Fickian yet non-Gaussian diffusion reported in heterogeneous
diffusion phenomena in various soft matter and biological systems.

To explain the non-Gaussian diffusion and its crossover to a Gaussian
behavior, we start by considering the diffusion of a single tracer. For time
intervals shorter than $\langle\tau^\mathrm{S}\rangle$, the sampled diffusivity
$D_\mathrm{S}$ of the tracer remains almost constant. On this time scale, the
single tracer diffusion is normal with the Gaussian propagator
\begin{eqnarray}
\label{eq:propagator}
G(x,\mathit{\Delta}|D_\mathrm{S})=\frac{1}{\sqrt{4\pi D_\mathrm{S}\mathit
\Delta}}\exp\left(-\frac{x^2}{4D_\mathrm{S}\mathit\Delta}\right),
\mathit{\Delta}\lesssim\langle\tau^\mathrm{S}\rangle.
\end{eqnarray}
When $\mathit\Delta\gg\langle\tau^\mathrm{S}\rangle$, the above Gaussian
limit no longer holds as the tracer explores the spatiotemporally varying
local diffusivity field. Conversely, on time scales at which the Gaussian
approximation \eqref{eq:propagator} holds, the van-Hove self-correlation
function $p(x_{\mathit\Delta},\mathit\Delta)$ can be evaluated via a
superstatistical approach \cite{beck2003superstatistics,chechkin2017,
sposini2018}. Using the PDF $\psi_\mathrm{S}$ from
Eq.~\eqref{eq:sampled_D_dist} of the sampled diffusivity, it can be
simply written as 
\begin{eqnarray}
\label{eq:superstatistics}
p(x_{\mathit\Delta},\mathit\Delta)=\int_0^\infty\mathrm{d}D_\mathrm{S}
G(x_{\mathit\Delta},\mathit\Delta|D_\mathrm{S})\psi_\mathrm{S}(D_\mathrm{S}),
\end{eqnarray}
where $x_\mathit{\Delta}\lesssim r_\mathrm{c}$ and $\mathit\Delta\lesssim
\langle\tau^\mathrm{S}\rangle$.
Performing this superstatistical integral above, we obtain the analytic
expression for the Van-Hove self-correlation function in the form
\begin{eqnarray}
\nonumber
p(x_{\mathit\Delta},\mathit\Delta)&\propto&\sqrt{\frac{\pi}{4D_\mathrm{env}
\mathit\Delta}}\mathrm{erfc}\left(\frac{|x_{\mathit\Delta}|}{2\sqrt{D_\mathrm{
env}\mathit\Delta}}+\sqrt{\frac{D_\mathrm{env}}{D_0}}\right)\\
&&\times\exp\left(\frac{x_{\mathit\Delta}^2}{4D_\mathrm{env}\mathit\Delta}
+\frac{D_\mathrm{env}}{D_0}\right),
\label{eq:exact}
\end{eqnarray}
where $\mathrm{erfc}(y)=(2/\sqrt{\pi})\int_y^{\infty}\exp(-t^2)\mathrm{d}t$
is the complementary error function. The derivation of Eq.~\eqref{eq:exact}
is detailed in App.~\ref{sec:appendixC}. In Figs.~\ref{fig3}(a) and (b), the
above theoretical approximation is shown as the dashed line along with the
simulations data. At short time $\mathit\Delta=1$, for which the superstatistics
is valid, the theoretical curve successfully explains the data, corresponding
to $|x_{\mathit\Delta}|\lesssim r_c$. This expression deviates from the
data, as expected, when $|x_{\mathit\Delta}|\gtrsim r_c$ in the tail part as
the particles traverse multiple extreme basins beyond the correlation length
of the order of $r_\mathrm{c}$.

We obtain further insight into the intricate behavior of the Van-Hove
self-correlation function \eqref{eq:exact} by simplifying it under the
condition $\mathit\Delta\ll|x_{\mathit\Delta}|^2D_\mathrm{env}^{-1}$, which
yields 
\begin{eqnarray}
\label{eq:asymptotic}
p(x_{\mathit\Delta},\mathit\Delta)\approx\mathscr{N}\frac{\exp\left(-\frac{
|x_{\mathit\Delta}|}{\sqrt{D_0\mathit\Delta}}\right)}{|x_{\mathit\Delta}|+
2D_\mathrm{env}\sqrt{\mathit\Delta/D_0}},
\end{eqnarray}
where $\mathscr{N}$ is the normalization factor. With this result, the
essential form of the van-Hove function is attained as follows: (1) in the
quenched limit, the environmental fluctuations are negligible ($D_\mathrm{
env}\approx0$) and $|x|\gg2D_\mathrm{env}\sqrt{\mathit\Delta/D_0}$. Hence,
the Van-Hove self-correlation function simplifies to
\begin{eqnarray}
\label{eq:asymptotic_quenched}
p(x_{\mathit\Delta},\mathit\Delta)\propto\frac{1}{|x_{\mathit\Delta}|}\exp
\left(-\frac{|x_{\mathit \Delta}|}{\sqrt{D_0\mathit\Delta}}\right),
\end{eqnarray}
reproducing the analytic form reported in Ref.~\cite{luo2019}. The prefactor
$|x_{\mathit \Delta}|^{-1}$ accounts for the cusp at the center observed in
the simulations results shown in Fig.~\ref{fig3}(a). (2) In the strong
annealing limit (i.e., $|x|\ll2D_\mathrm{env}\sqrt{\mathit\Delta/D_0}$),
when the environmental change is significantly larger than the particle
diffusion, the Van-Hove self-correlation function behaves as a Laplace
distribution,
\begin{eqnarray}
\label{eq:asymptotic_laplace}
p(x_{\mathit\Delta},\mathit\Delta)\propto\exp\left(-\frac{|x_{\mathit
\Delta}|}{\sqrt{D_0\mathit\Delta}}\right),
\end{eqnarray}
as seen in Fig.~\ref{fig3}(b).

The theoretical Van-Hove self-correlation functions \eqref{eq:exact} to
\eqref{eq:asymptotic_laplace} are valid when the superstatistical average can
be applied, on time scales $<\tau^S$. The exponential-like van-Hove function
changes into the Gaussian profile at times much longer than the homogenization
time $\tau_c$, see the plots of $p(x_{\mathit\Delta},\mathit\Delta)$ at $\Delta
=10,000$ in Figs.~\ref{fig3}(c) and (d). Here, the tracer samples a large number
of instantaneous diffusivities over a large distance, crossing a correlation
time analogously to the crossover to Gaussian behavior in the fluctuating
diffusivity model. The tail of $p(x_{\mathit\Delta},\mathit\Delta)$ then
gradually approaches a Gaussian PDF, in line with the central limit theorem.
In particular, for the annealed case (Fig.~\ref{fig3}(d)), the environmental
fluctuations assist the localized tracer in escaping from deep traps, which
results in a perfect Gaussian distribution, while in the quenched case a
cusp persists at the origin (Fig.~\ref{fig3}(c)).

\subsection{Non-Gaussianity Parameter}

We continue to examine the deviations from Gaussianity of the spatiotemporally
heterogeneous diffusion process on the disordered energy landscape in terms of
the non-Gaussianity parameter (i.e., the excessive kurtosis), for a centered
PDF,
\begin{equation}
\mathrm{NGP}=\frac{1}{3}\mathrm{Kurt}[x]-1=\frac{1}{3}\frac{\langle x^4\rangle}{
\langle x^2\rangle^2}-1
\end{equation}
where Kurt is the kurtosis~\cite{metzler2014}, and $x$ is one of the Cartesian components in $\mathbf{r}$. Due to the finite-observation
time effects shown in the tail part of the Van-Hove self-correlation functions,
%\textcolor{blue}{(which part of the discussion do you refer to?)}
we here employ the modified $\mathrm{NGP}$ (mNGP)
\begin{eqnarray}
\label{eq:NGP}
\widehat{\mathrm{NGP}}=\mathrm{NGP}-\mathrm{NGP}_0.
\end{eqnarray}
Here $\mathrm{NGP}_0$ represents the NGP for a truncated normal distribution
defined on the interval $[-\sqrt{2\log(N_\mathrm{tr}T/\mathit\Delta)},\sqrt{2
\log(N_\mathrm{tr}T/\mathit\Delta)}]$ in which $N_\mathrm{tr}$ denotes the
number of sample trajectories used in the analysis. The validity of $\widehat{
\mathrm{NGP}}$ for the analysis of non-Gaussianity in our system is
demonstrated in detail in App.~\ref{sec:appendixD}.

\begin{figure}
\includegraphics[width=0.37\textwidth]{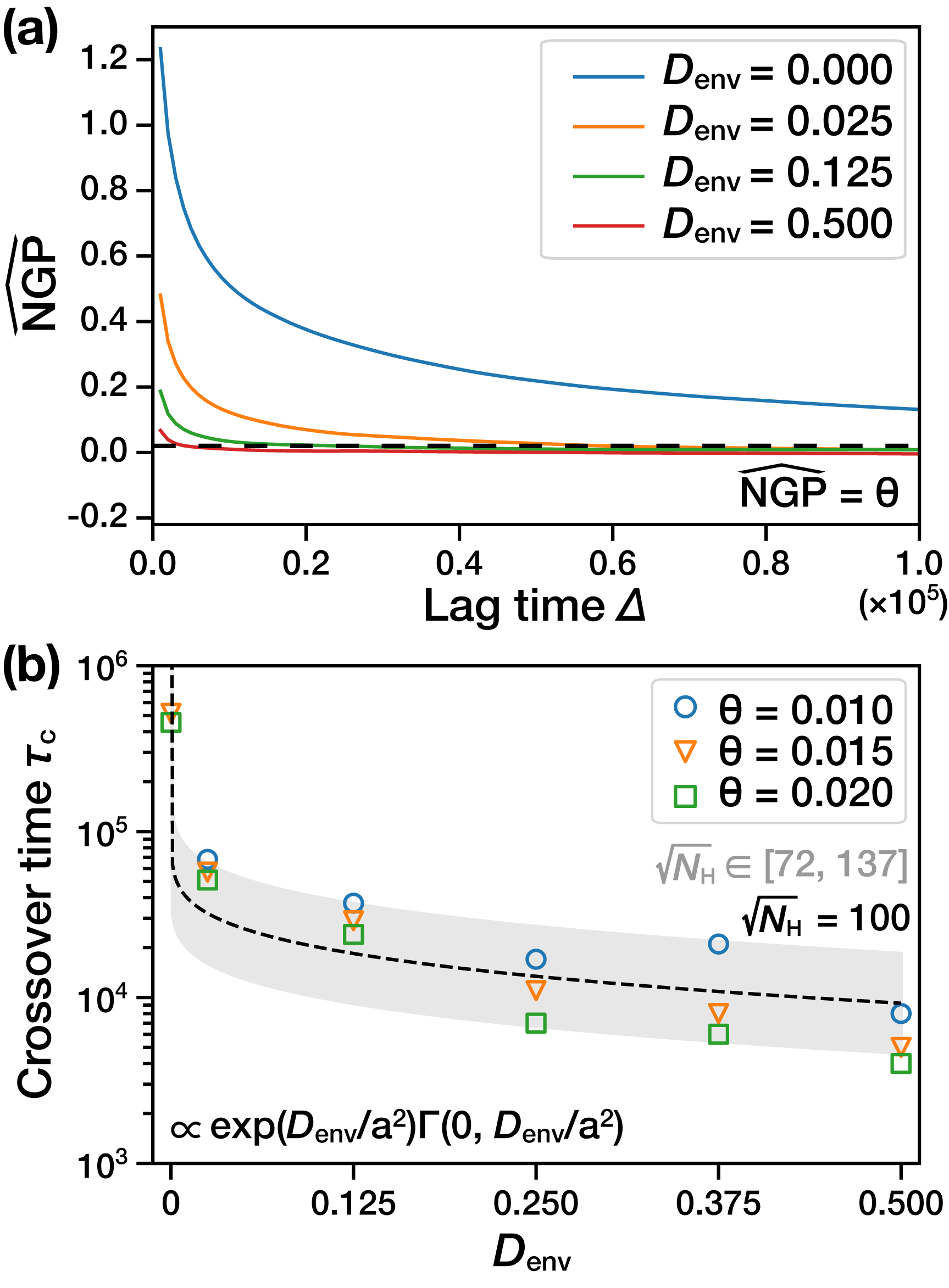} 
\caption{\textbf{Estimation of the non-Gaussianity and the crossover time.} (a)
$\widehat{\mathrm{NGP}}$ [Eq.~\eqref{eq:NGP}] as function of the lag time
$\mathit\Delta$ for several values of $D_\mathrm{env}$. Each curve was obtained
from $10^3$ simulated trajectories of length $T=2\times10^6$. The horizontal
dashed line depicts the threshold value $\theta=0.025$, see Eq.~\eqref{eq:tauc}.
(b) Crossover time $\tau_\mathrm{c}$ [Eq.~\eqref{eq:tauc}] against $D_\mathrm{
env}$ for three threshold values $\theta$. The gray shade indicates the region
of theoretical homogenization times [Eq.~\eqref{eq:crosstime}] with $\sqrt{N_
\mathrm{H}}\in[72,137]$. The black dashed line depicts the theoretical formula
\eqref{eq:crosstime} for the homogenization time, for $\sqrt{N_\mathrm{H}}=100$.}
\label{fig5}
\end{figure}

Figure \ref{fig5}(a) shows the variation of $\widehat{\mathrm{NGP}}$ with lag
time $\mathit\Delta$ from simulated trajectories for varying $D_\mathrm{env}$.
It demonstrates that for all cases the tracers perform non-Gaussian diffusion
at short times with a positive value of $\widehat{\mathrm{NGP}}$. The mNGP
gradually decays towards zero as $\mathit\Delta$ increases. The tracer motion
tends to recover Gaussianity faster with increasing $D_\mathrm{env}$, as here
the environmental fluctuations help the immobile tracer to escape from the
trapped site.

We define the crossover time $\tau_\mathrm{c}$ at which the short-time
non-Gaussianity crosses over to a Gaussian diffusion,
\begin{eqnarray}
\label{eq:tauc}
\tau_\mathrm{c}=\min\left\{\mathit{\Delta}|\widehat{\mathrm{NGP}}=\theta
\right\},
\end{eqnarray}
as estimated by the threshold value $\theta$ (corresponding to the dashed
line in Fig.~\ref{fig5}(a) for $\theta=0.025$). That is, the time scale
$\tau_\mathrm{c}$ refers to the moment when the NGP reaches the threshold
value $\theta$ for the first time. In Fig.~\ref{fig5}(b), we plot $\tau_
\mathrm{c}$ against $D_\mathrm{env}$ for three values of $\theta$. We
observe that regardless of the threshold values the estimated value of
$\tau_\mathrm{c}$ displays consistent variations with $D_\mathrm{env}$.
Namely, the crossover times gradually decrease with $D_\mathrm{env}$ in the
range from $10^4$ to $10^5$. It turns out that the estimated values are in
good agreement with the homogenization time that the Gaussian distribution
is recovered in the Van-Hove self-correlation function $p(x_{\mathit\Delta},
\mathit\Delta)$. For instance, the annealed system with $D_\mathrm{env}=0.5$
is found to have $\tau_c\approx10^4$, which is in good agreement with the
value in Fig.~\ref{fig3}(d) in which $p(x_{\mathit\Delta},\mathit\Delta)$ has
a Gaussian profile at $\mathit\Delta=10^4$. In the following Section, we
present the concept of homogenization and an analytic theory accounting for
the timescale $\tau_c$ in terms of the relevant system parameters of the
annealed extreme landscape.

\begin{figure}
\includegraphics[width=0.35\textwidth]{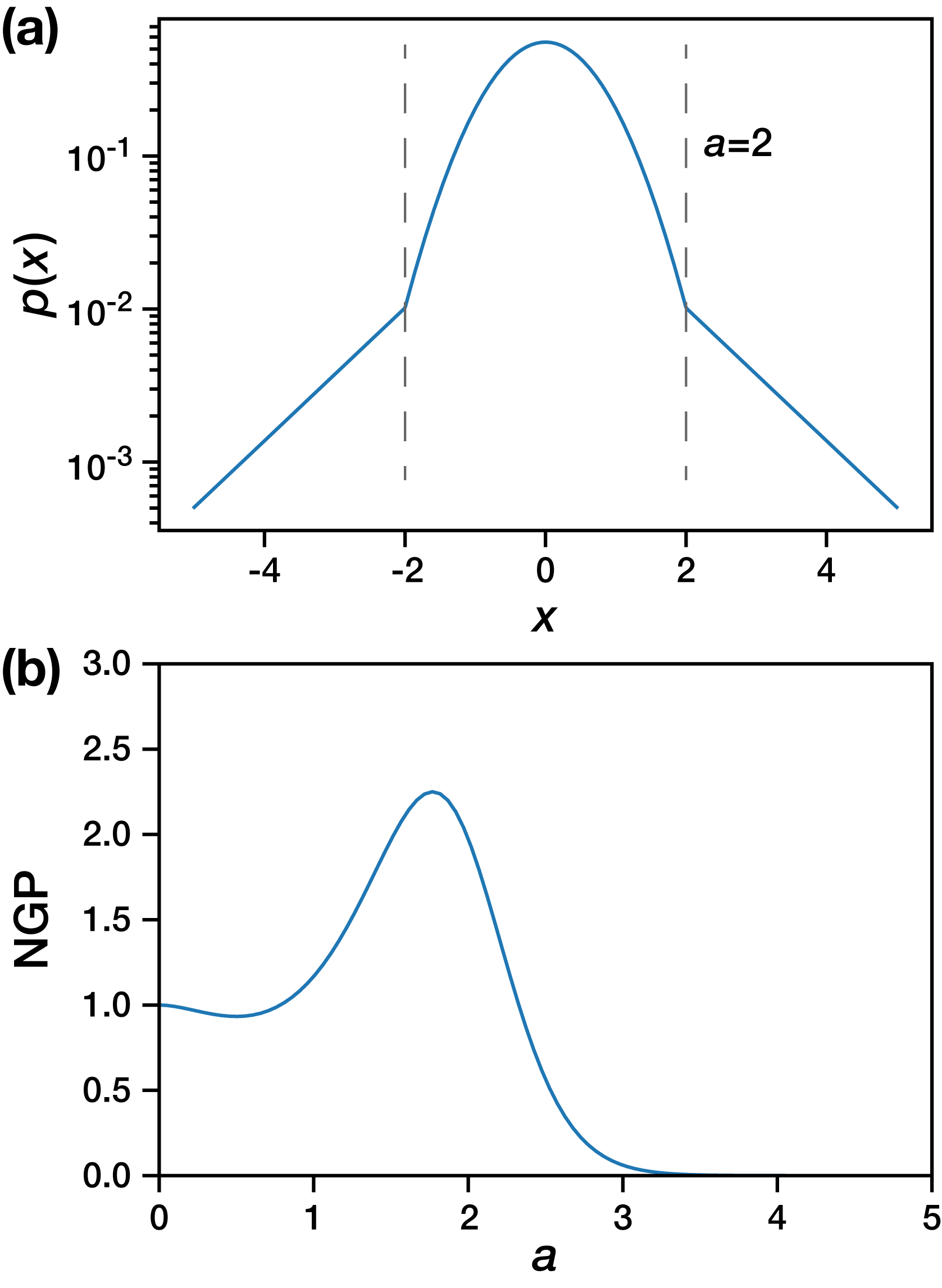}
\caption{(a) The composite PDF (\ref{combine}) with a Gaussian core and far Laplace tails for $a=2$. (b) Corresponding NGP (i.e., the excessive kurtosis) as a function of the width $a$ of the central Gaussian core. Already for $a\approx3$, NGP is approximately zero, as the weight of the Laplace tails becomes negligible. }
\label{fig_ngp}
\end{figure}

We stop to note that a zero value of the non-Gaussianity parameter does not
necessarily mean that the far tails of a PDF are indeed Gaussian. Namely, if
the PDF has a dominant central Gaussian component, the contributions to the
non-Gaussianity parameter of far non-Gaussian tails may simply become
negligible, despite being present in the PDF. This can be shown, e.g., by
calculating NGP for the composite PDF \cite{wei2025}
\begin{equation}
\label{combine}
p(x)=\mathscr{N}\left[e^{-a+a^2-x^2}\theta(a-|x|)+e^{-|x|}\theta(|x|-a)\right],
\end{equation}
in which $\mathscr{N}$ is the normalization constant and $\theta(x)$ is the
Heaviside step function. At $a$, that is, exponential tails are connected to
a central Gaussian. The composite PDF (\ref{combine}) along with the NGP is
shown in Fig.~\ref{fig_ngp}. We see that for $a=0$, i.e., a single Laplace
PDF, the limiting value $\mathrm{NPG=1}$ is reached, transiently increases for
growing $a$, until it finally reaches the value $\mathrm{NPG=0}$. This value
is approximately realized at $a=3$, despite the far Laplace tails.

\section{Homogenization of the environment and ergodicity}
\label{sec:homogenization}

As investigated in the previous Section, the short time non-Gaussian diffusion
in the heterogeneous extreme landscape eventually approaches an effective
Gaussian shape over time. This crossover from a non-Gaussian to a Gaussian
PDF in the long-time regime is termed \textit{homogenization}. We now focus
on this homogenization process, before addressing the ergodicity breaking
parameter of the dynamics.

\subsection{Homogenization}
\label{subsec:homogenization}

One can anticipate that in an annealed random environment the tracer particle
initially is determined by the spatiotemporally heterogeneous diffusion. Once,
however, the tracer has explored a sufficient fraction of distinct sites of a
given diffusivity field, it has sampled a substantial fraction of the given
random environment and thus reaches an effectively homogeneous Gaussian PDF
with a mean diffusivity. This is what we refer to as the homogeneous process.
If the diffusivity field is self-similar such that a part of the entire field
mirrors the whole random diffusivity landscape, we may expect that sampling
this partial diffusivity field suffices for a tracer particle to achieve
homogenization.

\begin{figure}
\includegraphics[width=0.37
\textwidth]{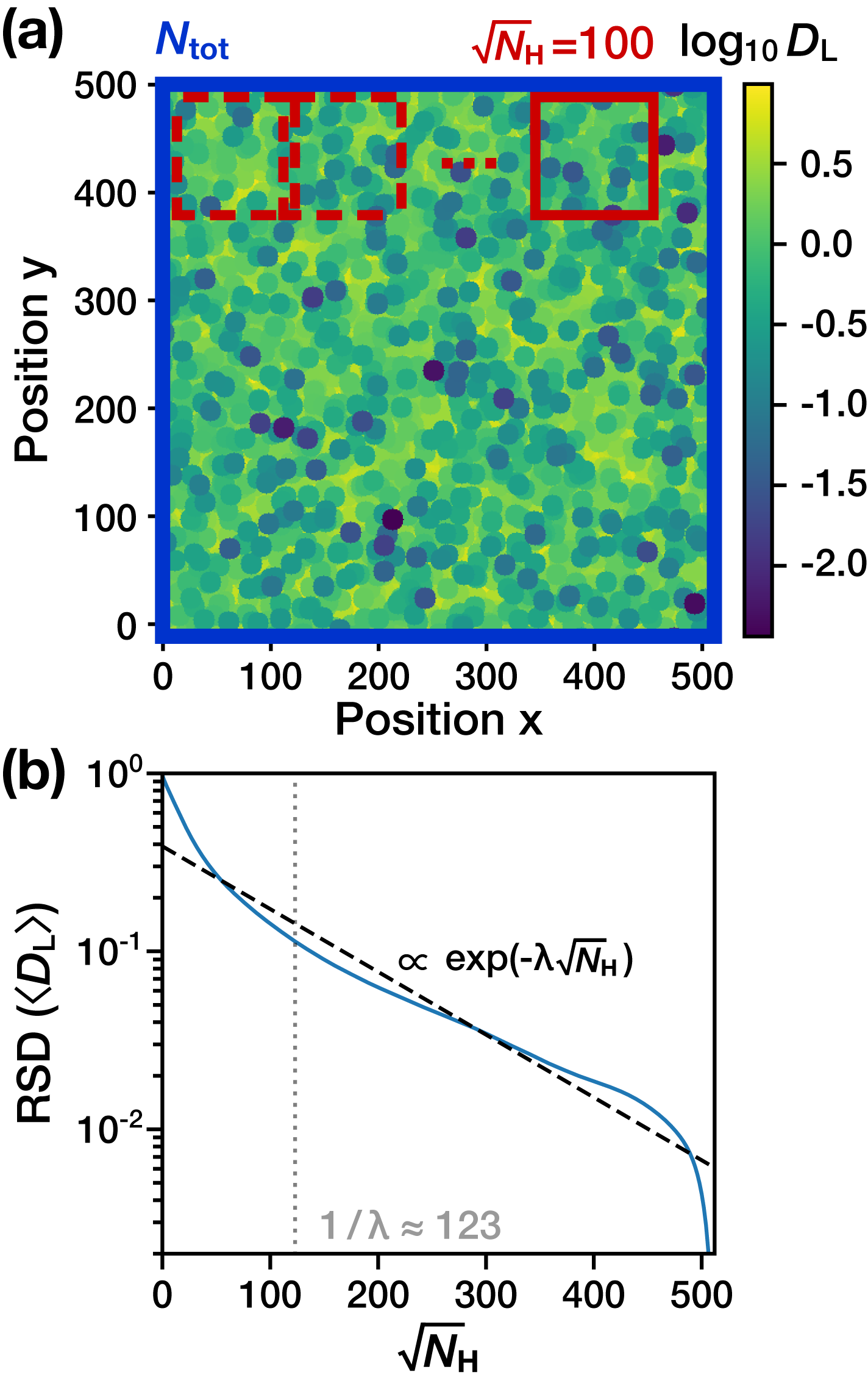} 
\caption{\textbf{Analysis of the self-similarity of the extreme landscape.}
(a) Simulated local diffusivity field (blue box) with samples of a subset
field of size $N_\mathrm{H}$ (red box). Using the sliding window (depicted
as the dashed boxes in the plot) with periodic boundary condition, a total
of $L^2=512^{2}$ subsets are defined. (b) Relative standard deviation (RSD)
of the mean local diffusivities, Eq.~\eqref{eq:rsd}, as a function of the
window size $N_\mathrm{H}$ (solid line). The dashed line represents an
exponential decay $\propto\exp(-\lambda\sqrt{N_H})$ with a best fit value
$\lambda=0.0081$. The dotted line denotes the characteristic constant
$1/\lambda\approx123$ for $\sqrt{N_H}$. See text for details.}
\label{fig7}
\end{figure}

As an example to demonstrate this idea, we present in Fig.~\ref{fig7}(a) a
simulated diffusivity field $\{D_\mathrm{L}\}$ of size $N_\mathrm{tot}$
(blue box,  $N_\mathrm{tot}=512^2$)
with several subsets of size $N_\mathrm{H}$ depicted as the red boxes. The
local diffusivity field in the subset is defined as a square box with a side
length of $\sqrt{N_\mathrm{H}}$. 
%\textcolor{blue}{Why is then $\sqrt{N_H}$ indicated in the figure? Needs clarification!} 
Distinct subsets are created
using a sliding window with the periodic boundary condition (the dashed boxes
in Fig.~\ref{fig7}(a)), resulting in a total of $L^2$ sliding windows. For
each subset diffusivity field, we measure the mean local diffusivity $\langle
D_\mathrm{L}\rangle$ within the window. If a subset of size $N_\mathrm{H}$
is a self-similar representation of the entire environment, the mean local
diffusivities from distinct subsets will not differ much from each other,
i.e., exhibiting a small deviation from the diffusivity of the entire field.
However, if the system is not self-similar or the subset size is not
sufficiently large, the system will display substantial deviations in their
mean local diffusivities.

We define the relative standard deviation (RSD) of the mean local diffusivities 
\begin{eqnarray}
\label{eq:rsd}
\mathrm{RSD}(\langle D_\mathrm{L}\rangle)=\frac{\sqrt{\mathbb{E}[\langle D_
\mathrm{L}\rangle^2]-\mathbb{E}[\langle D_\mathrm{L}\rangle ]^2}}{\mathbb{
E}[\langle D_\mathrm{L}\rangle]},
\end{eqnarray}
where $\mathbb{E}[\cdots]$ denotes ensemble-averaging over the distinct sliding
windows. Figure~\ref{fig7}(b) shows a numerical estimate for RSD as a function
of $\sqrt{N_\mathrm{H}}$ (solid line). The RSD monotonically decays with the
size of the subset diffusivity field, supporting the idea of the discussed
self-similarity property. The result shows that the decrease of RSD tends to
follow an exponential decay with $\sqrt{N_\mathrm{H}}$. Fitting with an
exponential function $\propto\exp(-\lambda\sqrt{N_\mathrm{H}})$ (dashed line)
reveals that $1/\lambda\approx 123$ is the characteristic size of a subset
field to ensure sufficient self-similarity. Namely, if the local subset field explored by a tracer is larger than the characteristic size, the RSD exponentially decreases to zero, indicating that the tracer has almost the same mean diffusivity regardless of the position of the sampling subset field.

We now conceive a theoretical description for the homogenization time based
on the idea of the self-similar local field. Let $N_\mathrm{tot}=L^2$ be the
total number of lattice sites of our diffusivity field and $N_\mathrm{H}$ be
the number of distinct lattice sites visited by the tracer as required for
achieving homogenization. As argued above, the local sites included in the
field $N_\mathrm{H}$ are sufficient to capture the essential features of the
spatiotemporally heterogeneous landscape of the entire field. We also define
$M$ as the number of steps required for a tracer to visit $N$ distinct sites
during its random walk. On a 2D lattice, the average number of distinct
sites $N$ visited by $M$ random steps is known to be \cite{biroli2022number}
\begin{eqnarray}
\label{eq:relation}
N\approx\frac{\pi M}{\log(8M)}.
\end{eqnarray}
Based on this expression, we derive the average number of visits $\gamma$ to
every distinct site after $M$ random steps as a function of $N$,
\begin{eqnarray}
\label{eq:gamma}
\gamma(N)\equiv\frac{M}{N}=-\frac{1}{\pi}\mathrm{W}\left(-\frac{\pi}{8N}\right),
\end{eqnarray}
where $\mathrm{W}(x)$ denotes the Lambert W-function~{\color{blue}\cite{corless1996lambert}}.
From this relation, the time $\tau_\mathrm{c}$ required to sample $N_\mathrm{
H}$ distinct sites can be obtained on the mean-field level as 
\begin{eqnarray}
\nonumber
\tau_\mathrm{c}&\approx&\gamma(N_\mathrm{H})\frac{N_\mathrm{H}}{N_\mathrm{tot}}
\sum_{\mathbf{r}}^{N_\mathrm{tot}}\tau[\omega(\mathbf{r})]\\
&\approx&\gamma(N_\mathrm{H})\frac{N_\mathrm{H}}{N_\mathrm{tot}}\int_0^\infty
\mathrm{d}\omega\,\frac{N_\mathrm{tot}}{\omega+\omega_\mathrm{env}}f_\omega.
\end{eqnarray}
In the last expression the summation term is approximated by an integral, in
which $f_\omega=\exp(-\omega/4)/4$ is the density of states. Performing the
integral, we obtain the analytic expression for $\tau_c$ in terms of the
annealing rate ($D_\mathrm{env}$) and the size of a self-similar subset field
in the form
\begin{equation}
\label{eq:crosstime}
\tau_\mathrm{c}=\frac{\gamma(N_\mathrm{H})N_\mathrm{H}}{4}\exp\left(\frac{
D_\mathrm{env}}{a^2}\right)\Gamma\left(0,\frac{D_\mathrm{env}}{a^2}\right).
\end{equation}
Here, $\Gamma(s,x)$ is the upper
incomplete gamma function explained above in Eq.~\eqref{eq:mean_sampled_diffusivity}.  

We validate our analytical result \eqref{eq:crosstime} with stochastic
simulations. Fig.~\ref{fig5}(b) shows the simulated crossover times at
which the initial non-Gaussian diffusion is crossing over to Gaussian
dynamics in terms of the mNGP, with the criterion \eqref{eq:tauc}, for
various annealed extreme landscapes. These cross-over times are fitted with
Eq.~\eqref{eq:crosstime} in which $N_\mathrm{H}$ is a free parameter. We see
that our analytic results reproduces nicely the measured crossover times with
$\sqrt{N_\mathrm{H}}\approx100$ (dashed line). The gray shade region depicts
the range of the subset size, $\sqrt{N_\mathrm{H}}\in[72,137]$, that is used
for simulating the cross-over times $\tau_c$ against $D_\mathrm{env}$ for
different thresholds $\theta$.

\subsection{Ergodicity breaking parameter}
\label{sec:EB}

\begin{figure}
\includegraphics[width=0.46\textwidth]{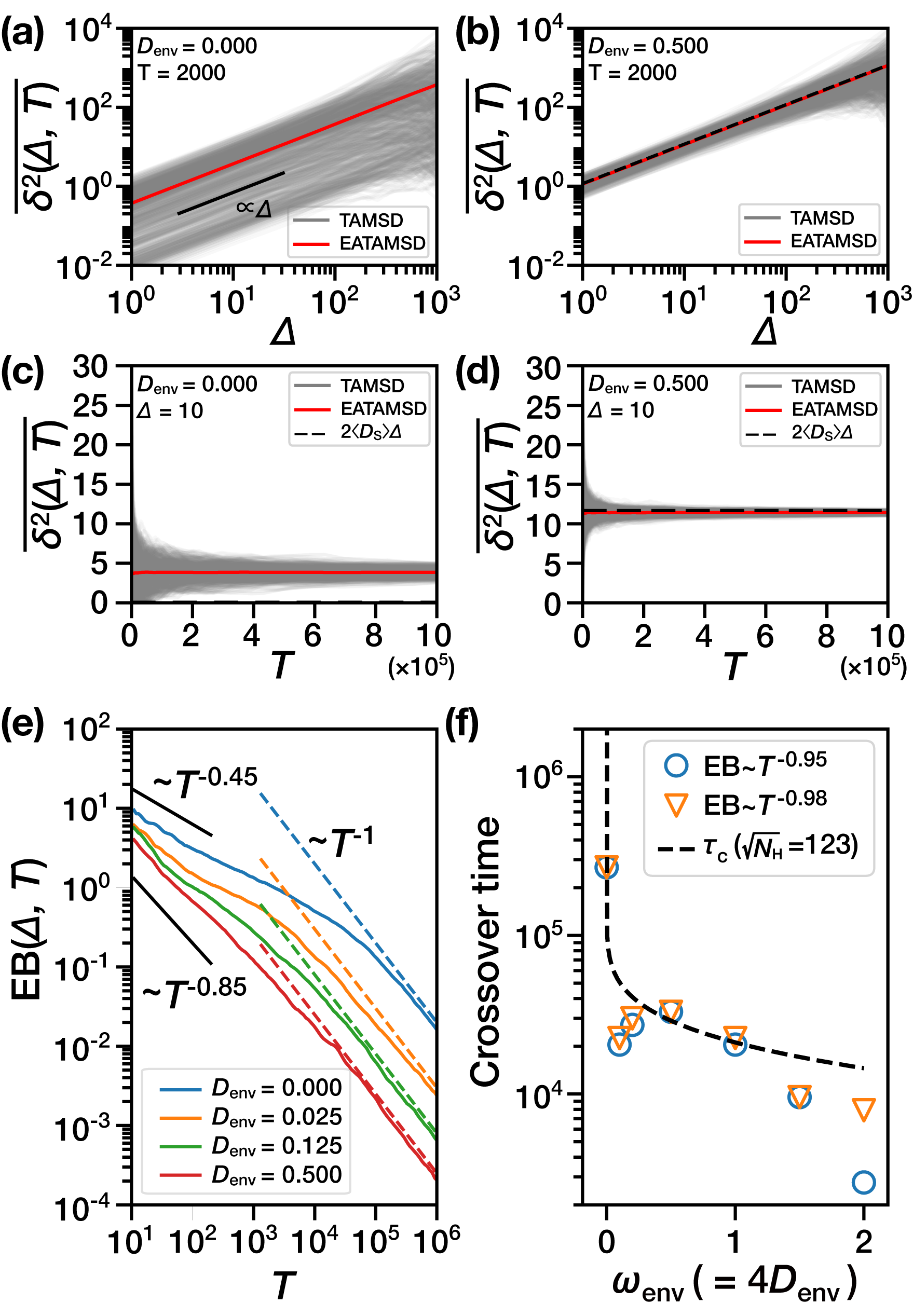}
\caption{\textbf{Particle-to-particle heterogeneity and ergodicity
breaking parameter.} (a, b): Individual TAMSDs (gray) as a function of lag
time $\mathit\Delta$ and corresponding ensemble-average (EATAMSD, red). (a)
$D_\mathrm{env}=0$ (quenched). (b) $D_\mathrm{env}=0.5$ (annealed). (c, d)
TAMSDs as a function of $T$ at $\mathit{\Delta}=10$. (c) $D_\mathrm{env}
=0$ (quenched). (d) $D_\mathrm{env}=0.5$ (annealed). (e) EB parameter
[Eq.~\eqref{eq:eb}] as a function of $T$ for various values of $D_\mathrm{
env}$. The solid black lines depict the two empirical power-laws $T^{-0.45}$
and $T^{-0.85}$, the dashed lines represent the power-law $T^{-1}$. (f)
Crossover times beyond which the EB parameter begins to decay as $\simeq
T^{-\theta_\mathrm{EB}}$, with $\theta_\mathrm{EB}=0.95$ (blue) and 0.98
(orange). The dashed line indicates the theoretical homogenization time
$\tau_\mathrm{c}$ [Eq.~\eqref{eq:crosstime}] with $\sqrt{N_\mathrm{H}}=100$.}
\label{fig8}
\end{figure}

In addition to the temporal heterogeneity from a single trajectory, we now
investigate another inherent form of heterogeneity, i.e., the heterogeneity
from one trajectory to another \cite{sposini2018}. In the preceding Section,
we introduced the homogenization time $\tau_\mathrm{c}$ accounting for the
time scale at which the temporal heterogeneity is effectively averaged out.
Beyond this timescale, the tracers exhibit a homogeneous Gaussian diffusion
pattern. In the regime $t\lesssim\tau_\mathrm{c}$, in contrast, individual
particles display pronounced trajectory-to-trajectory heterogeneity, due to
the distinct explored local energy landscape.

Figs.~\ref{fig8}(a), (b) show TAMSDs versus the lag time for simulated
individual trajectories when $D_\mathrm{env}=0$ and $=0.5$. In both panels,
EATAMSD is also shown (red lines). On the one hand, a substantial heterogeneity
is visible between individual trajectories when the environment is quenched
($D_\mathrm{env}=0$) and the observation time $T(=2000)$ is much shorter than
the homogenization time $\tau_c$. On the other hand, the annealing of the
random landscape may significantly reduce the trajectory-to-trajectory
heterogeneity even for short $T$, as seen in Fig.~\ref{fig8}(b). In
Figs.~\ref{fig8}(c), (d), we display the TAMSDs against $T$ for a given
$\mathit\Delta$. As expected, the variation in the TAMSDs is larger in the
quenched case as compared to the annealed case, and generally the TAMSD
amplitude variations tend to be reduced as $T$ is larger. Particularly, in
the annealed random landscape (Fig.~\ref{fig8}(d)), individual TA MSDs
converge to the MSD, Eq.~\eqref{eq:fickianlaw} (dashed line), when $T$
exceeds the homogenization time $\tau_\mathrm{c}\approx10^4$.

We analyze the trajectory-to-trajectory heterogeneity in terms of the
ergodicity breaking (EB) parameter \cite{he,deng,metzler2014,pt,
jeon2016protein}
\begin{eqnarray}
\label{eq:eb}
\mathrm{EB}(\mathit\Delta,T)=\frac{\left<\left[\overline{\delta^2(\mathit
\Delta,T)}\right]^2\right>-\left<\overline{\delta^2(\mathit\Delta,T)}\right>^2
}{\left<\overline{\delta^2(\mathit\Delta,T)}\right>^2}.
\end{eqnarray}
In Fig.~\ref{fig8}(e), we evaluate $\mathrm{EB}$ as a function of $T$ for
various values of $D_\mathrm{env}$. The EB parameter monotonically decreases
as $D_\mathrm{env}$ increases, suggesting that the annealing environment
indeed tends to reduce the particle-to-particle heterogeneity, as discussed
above. Notably, EB decays with two distinct power-law scalings: That is, a
phenomenological power-law with an exponent less than unity for $t<\tau_c$
and the typical power-law $T^{-1}$ for Gaussian diffusion after homogenization
\cite{metzler2014}. The first, anomalous regime is attributed to the
spatiotemporally heterogeneous diffusion in the annealed random geometries. The strongest heterogeneity occurs in the quenched case,
for which the power-exponent is smallest ($\approx0.45$). As the annealing
rate increases, the heterogeneous effect is weakened and, accordingly, the
power-law exponent becomes larger. Beyond the homogenization timescale, the
EB parameter converges to the universal relation $\mathrm{EB}\sim T^{-1}$
satisfied by typical homogeneous ergodic diffusion processes, such as
Brownian motion or the fractional Langevin equation model \cite{metzler2014,
deng,pt,he,jeon2016protein}.

We estimate the crossover time at which the EB parameter starts to decay as
$\simeq T^{-1}$. Because $T^{-1}$ is the expected asymptotic scaling, we
first define a threshold exponent $\theta_\mathrm{EB}$ in the power-law
scaling of the EB parameter $T^{-\theta_\mathrm{EB}}$ and find the
instant when the slope of the EB parameter, for the first time, becomes
$\theta_\mathrm{EB}$. For two threshold exponents $\theta_\mathrm{EB}=0.95$
and $=0.98$, we estimate the crossover timescales (symbols) for various values
of the annealing rate [Fig.~\ref{fig8}(f)]. The data are compared to the
behavior of the homogenization time $\tau_c$ [Eq.~\eqref{eq:crosstime},
$\sqrt{N_\mathrm{H}}=100$] shown in Fig.~\ref{fig5}(b). Excellent agreement
between the two quantities is found, demonstrating that the convergence to
ergodicity at long times is intimately related to the homogenization process.

\section{Discussion and conclusions}
\label{sec:discussion}

We here conceived an annealed extreme landscape model and performed a
comprehensive computational and theoretical investigation on the non-Gaussian
properties of a particle diffusing on this annealed heterogeneous environment.
The main focus of this study is to characterize Fickian yet non-Gaussian
diffusion in an annealed heterogeneous environment, with particular emphasis
on how it relates to the annealing rate of the system and its convergence to
homogeneous diffusion.

The four key findings of our study are as follows:

(1) Fickian diffusion is universal, regardless of the system's annealing
rate, provided that the initial condition is stationary. We theoretically
obtained the Fickian diffusion relation, $\langle\Delta x(t)^2\rangle
\approx2\langle D_\mathrm{S}\rangle t$, which successfully explained the
MSDs computed from the simulations. In this relation, the system's annealing
increases the mean sampled diffusivity, $\langle D_\mathrm{S}\rangle$,
by assisting particles to escape from regions of slow diffusivity.

(2) The annealing of the system impacts the non-Gaussianity. When the
annealing is slow, the Van-Hove self-correlation function is highly
non-Gaussian, showing a pronounced peak around $\Delta x=0$, as particles
spend more time in regions of low diffusivity. When the annealing is faster,
particles readily escape from slow diffusivity regions, resulting in a
reduced peak around $\Delta x=0$ in the Van-Hove self-correlation function.
The Van-Hove self-correlation functions for varying $D_\mathrm{env}$
computed from the simulated trajectories are explained via the concept of
superstatistics at short lag times and small $\Delta x~(<r_\mathrm{c})$.

(3) The annealing of the system reduces the homogenization time after which
the initial non-Gaussian particle diffusion crosses over towards a Gaussian
behavior. We found that this crossover emerges after the particle explores the
space to a degree that the local random environment is sufficiently
self-similar to the overall random landscape. As the particle samples the
random environment faster in a rapidly annealing medium, the crossover
timescales decrease monotonically with increasing $D_\mathrm{env}$.

(4) Finally, the annealing of the random environment reduces the
trajectory-trajectory heterogeneity. Before homogenization, each particle
experiences a different surrounding environment. However, beyond the
homogenization timescale, every particle effectively samples a sufficiently
similar random environment. Consequently, the diffusion dynamics of different
particles deviate negligibly from each other. This fact is corroborated by the
occurrence of the inverse power-law scaling in the ergodicity breaking
parameter, $\mathrm{EB}(T)\sim T^{-1}$, at times beyond the homogenization
time.

Let us now briefly discuss our findings with respect to previous studies on
Fickian yet non-Gaussian diffusion phenomena. The main origin for Fickian yet
non-Gaussian diffusion is the variability in the tracer diffusivity. The
varying diffusivity is often due to the heterogeneous environment in which
the tracer particles are embedded. Examples include the dynamics of beads
diffusing along lipid tubes \cite{wang2009anomalous} or in polymer networks
\cite{wang2009anomalous,wang2012brownian,xue2016probing}, and the motion of
tracers in crowded \cite{ghosh2015non,jeon2016protein,he_nc} or disordered
media \cite{chakraborty2020disorder}. As tracer particles explore such
environments, their diffusivity fluctuates spatiotemporally. This gives rise
to a varying diffusivity characterized by a diffusivity distribution $\psi(
D)$ \cite{wang2009anomalous,wang2012brownian,chechkin2017,sposini2018}, which
in turn produces a non-Gaussian van-Hove self-correlation function, as
discussed in Sec.~\ref{sec:3b}.

In most theoretical studies, the variability of the diffusivity is typically
modeled as a stochastic process, such as the square of the Ornstein-Uhlenbeck
process \cite{chechkin2017,sposini2018,sposini2024being,sposini2024being2,
jain2016diffusing,tyagi}, a barometric formula-like scenario
\cite{chubynsky2014diffusing} the Feller process~\cite{lanoiselee2018,
lanoiselee2018model}, or switching diffusivity models
\cite{miyaguchi2019brownian,doerries2022apparent,grebenkov2019unifying,
sabri2020elucidating}. These studies commonly demonstrate that the particle
diffusion in complex media leads to transient non-Gaussian diffusion over
short timescales, which eventually crosses over to Gaussian diffusion beyond
a characteristic correlation time, as predicted by the central limit theorem.
Additionally, Fickian diffusion dynamics $\langle\Delta x^2(t)\rangle\propto
t$ is observed when the initial ensemble is stationary. The Fickian yet
non-Gaussian diffusion observed in an annealed extreme landscape in this
study is in line with these previous findings. However, we stress that the
previous theoretical models that assume a random time-dependent diffusivity
do not fully account for the geometry and dynamics of the environment. In
contrast, our annealed extreme landscape model is based on a specific
realization of the geometry, represented by the local diffusivity field
$D_\mathrm{L}(\mathbf{r},t)$, which also incorporates the environmental
dynamics controlled by the annealing parameter $D_\mathrm{env}$. In this
setting, beyond the conventional discussions about a non-Gaussian van-Hove
self correlation function or the scaling of the MSDs, we succeeded in
relating the crossover timescale (i.e., the homogenization time) to the
energy landscape and the annealing dynamics of the environment. Moreover,
the dependence of the particle mean sampled diffusivity, $\langle D_\mathrm{
S}\rangle$, on the environmental annealing is a finding that has not been
considered or predicted in previous random diffusivity models
\cite{sposini2024being,sposini2024being2,lanoiselee2018,lanoiselee2018model}.

To conclude, a thorough understanding of a medium's energy landscape the
diffusive dynamics of a particle exploring the landscape are crucial to
accurately quantify heterogeneities in the particle diffusion, such as
non-Gaussianity and ergodicity breaking. Our study offers theoretical
insights into how these factors are linked to the key timescales governing
particle diffusion and demonstrates that the Fickian yet non-Gaussian
diffusion indeed emerges in an annealed heterogeneous media. We propose
several directions for future work related to our current study. Recently,
fluctuating diffusivity models have been studied on the topics of
diffusion-limited reactions and first-passage dynamics \cite{sposini2018first,
sposini2024being,sposini2024being2,lanoiselee2018,grebenkov2019unifying,
grebenkov2021exact}. As discussed, above, these models do not account
for annealing-dependent mean sampled diffusivity and typically assume an
\emph{immobile\/} reactant target, which conflicts the assumptions of an
annealing environment. A systematic understanding of diffusion-limited
reactions in dynamic heterogeneous media, based on the explicit simulation
of an annealed environment such as our annealed extreme landscape model
and comparisons with conventional time-dependent diffusivity models, is
needed. Furthermore, while our current model incorporates the environmental
spatiotemporal variation via a two-dimensional lateral diffusion with local
diffusivities $D_\mathrm{L}(\mathbf{r})$, many real systems may have more
complex geometry or higher dimension. Expanding the model to encompass these
additional complexities is a challenge for future research.

\begin{acknowledgments}
This work was supported by the National Research Foundation (NRF) of Korea
(grants RS-2023-00218927 \& RS-2024-00343900). We also acknowledge funding
from the German Science Foundation (DFG, grant ME 1535/13-2 and ME 1535/16-1).
\end{acknowledgments}

\appendix

\setcounter{figure}{0}

\renewcommand{\thefigure}{A\arabic{figure}}

\section{Simulations details}
\label{sec:appendixA}

Below we provide the details of the simulation algorithm, which involves five
input variables: $\omega_\mathrm{env}$, $r_c$, $L$, $\delta$, and $T$. These
represent the annealing rate of the environment, the critical radius of an
extreme basin, the size of the environment, the time resolution for perturbing
the environment, and the total simulation time, respectively. The position of a
particle at time $t$ is denoted by $\mathbf{x}(t)$. The particle random walk
is implemented using the Gillespie algorithm \cite{gillespie1977exact}, and
the auxiliary random field $U(\mathbf{r},t)$ is updated at every time interval
$\delta$, with $\delta\leq1$. The simulations output includes the time series
$\mathbf{x}(t)$ of the particle and the sampled diffusivity $D_\mathrm{S}(t)$.

\begin{algorithm}[H]
\caption{Diffusion on an annealed landscape}
\begin{algorithmic}
\Require{$\omega_{\mathrm{env}}$, $r_c$, $L$, $\delta$, $T$}
\State{$t \gets 0$}
\State{$\{ U(\mathbf{r},t)\} \gets $ auxiliary random field [Eq.~\eqref{eq:auxiliary}]}
\State{$\{V(\mathbf{r},t)\} \gets $ extreme landscape [Eq.~\eqref{eq:extreme_landscape}]}
\State{$\mathbf{x}(t) \gets $ initial position satisfying Eq.~\eqref{eq:stationary}}
\While{$t < T$}
    \State{$\mathbf{x}(t+1) \gets \mathbf{x}(t)$}
    \State{$\delta\text{sum} \gets 0$}
    \While{$\delta\text{sum} < 1$}
        \State{$\eta\text{sum} \gets 0$}
        \While{$\eta\text{sum} \leq \delta$}
            \State{$\omega \gets \omega_0 \exp[V(\mathbf{x},t+\delta\text{sum})]$}
            \State{$\eta \gets -\log [1-\text{unif}(0, 1)] / \omega$}
            \State{$\eta\text{sum} \gets \eta\text{sum} + \eta$}
            \If{$\eta\text{sum} \leq \delta$}
                \State{$\mathbf{x}(t+1) \gets$ one of its adjacent points}
            \EndIf
        \EndWhile
        \State{$\{ U(\mathbf{r},t+\delta\text{sum} + \delta) \} \gets$ disturbed $\{ U(\mathbf{r},t+\delta\text{sum})\}$}
        \State{$\{ V(\mathbf{r}, t+\delta\text{sum} + \delta) \} \gets$ extreme landscape}
        \State{$\delta\text{sum} \gets \delta\text{sum} + \delta$}
    \EndWhile
    \State{$t \gets t+1$}
    \State{$D_\text{S}(t) \gets \pi r_c^2 \exp \left [ V(\mathbf{x}(t), t)\right ]$}
\EndWhile
\Ensure{$\{\mathbf{x}(t), D_\text{S}(t) \mid t < T\}$}
\end{algorithmic}
\end{algorithm}

\section{PDFs from variable transformations}
\label{sec:appendixB}

Here, we derive the PDFs $f_\omega(\omega)$ of the escape rate of a particle
and of the sojourn time $f_\tau(\tau)$.

\subsection{PDF of escape rates}

The escape rate $\omega\in\mathbb{R}^+$ is defined by the transformation of
the potential $V\in\mathbb{R}$ as in Eq.~\eqref{eq:walkrate},
\begin{eqnarray}
\omega=\omega_0\exp(V).
\end{eqnarray}
The inverse transformation is given by
\begin{eqnarray}
V=\log(\omega/\omega_0),
\end{eqnarray}
which involves the Jacobian $|\mathrm{d}V/\mathrm{d}\omega|=\omega^{-1}$.
The PDF of $\omega$ is described as
\begin{eqnarray}
\nonumber
f_\omega(\omega)&=&\phi_V(V)\left|\frac{\mathrm{d}V}{\mathrm{d}\omega}\right|\\
\nonumber
&=&\frac{\exp[V-V_0-\exp(V-V_0)]}{\omega}\\
&=&\frac{t_0}{4}\exp\left(-\frac{\omega t_0}{4}\right)
\end{eqnarray}
with the unit simulation time $t_0=1$.

\subsection{PDF of sojourn times}

The sojourn time $\tau$ is defined by the transformation of the escape rate,
as in Eq.~\eqref{eq:tau},
\begin{eqnarray}
\tau=\frac{1}{\omega+\omega_\mathrm{env}}.
\end{eqnarray}
We note that $\tau$ satisfies $\tau\in(0,\omega_\mathrm{env}^{-1})$ due to the
constraint $\omega\in\mathbb{R}^+$. From the inverse transformation
\begin{eqnarray}
\omega=\frac{1}{\tau}-\omega_\mathrm{env}
\end{eqnarray}
we obtain the Jacobian $|\mathrm{d}\omega/\mathrm{d}\tau|=\tau^{-2}$. The
PDF of $\tau$ is then described as
\begin{eqnarray}
\nonumber
f_\tau(\tau)&=&f_\omega(\omega)\left|\frac{\mathrm{d}\omega}{\mathrm{d}\tau}
\right|\\
\nonumber
&=&\frac{t_0}{4}\exp\left(-\frac{\omega t_0}{4}\right)\frac{1}{\tau^2}\\
\nonumber
&=&\frac{t_0}{4\tau^2}\exp\left[-\frac{t_0}{4}\left(\frac{1}{\tau}-\omega_
\mathrm{env}\right)\right]\\
&=&\frac{\tau_0 \exp (\omega_\mathrm{env}\tau_0)}{\tau^2}\exp\left(-\frac{
\tau_0}{\tau}\right),
\end{eqnarray}
where $\tau_0=t_0/4$. In the quenched limit $\omega_\mathrm{env}=0$, we recover
the previous result obtained in the study of the quenched extreme landscape
\cite{luo2018}. Note that both the quenched and annealed systems have the same
power-law scaling $f_\tau(\tau)\sim\tau^{-2}$ for large $\tau$.

\section{MSD for a non-stationary diffusion trajectory}
\label{sec:nonstationary}

\begin{figure*}
\includegraphics[width=0.85\textwidth]{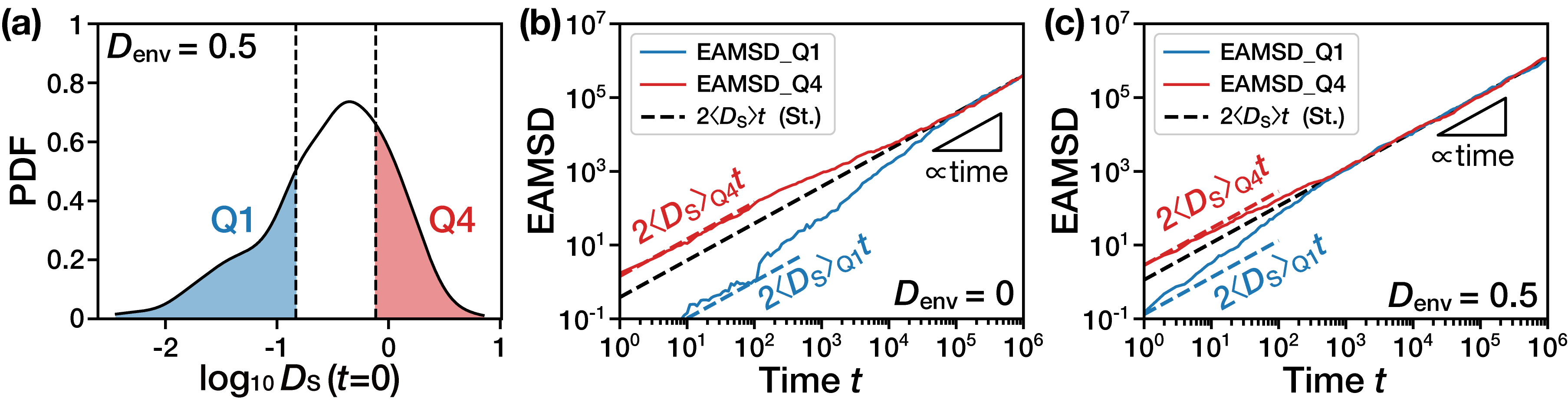} 
\caption{\textbf{MSD for a non-stationary ensemble.} (a) Sampled diffusivity
distribution for $D_\mathrm{env}=0.5$, estimated using the simulated
trajectories with stationary initial condition (black) and non-stationary
populations chosen from the slowest 25\% (Q1, blue) and the fastest 25\% (Q4,
red). (b) and (c) MSDs for non-stationary ensembles Q1 (blue) and Q4 (red) at
(b) $D_\mathrm{env}=0$ and at (c) $D_\mathrm{env}=0.5$. At short times, the
MSDs for the Q1 and Q4 groups exhibit the Fickian law $2\langle D_\mathrm{S}
\rangle_\mathrm{Q1}t$ and $2\langle D_\mathrm{S}\rangle_\mathrm{Q4}t$,
respectively. At long times, both MSDs converge to $2\langle D_\mathrm{S}
\rangle t$.}
\label{fig:nonstationary}
\end{figure*}

In the main text, we examined the MSD dynamics for stationary initial
conditions, in which particles are initially located according to the
stationary distribution \eqref{eq:stationary}. In Fig.~\ref{fig:nonstationary}(a),
we plot the PDF of the initial sampled diffusivities $D_\mathrm{S}(t=0)$ for
$D_\mathrm{env}=0.5$ (black solid line), estimated from 1,000 simulated
trajectories satisfying the stationary initial condition.

For a non-stationary initial condition, we selected the slowest 25\% (250
trajectories) of the stationary ensemble, referred to as Q1, and the
fastest 25\%, referred to as Q4, based on their initial sampled diffusivities.
This is graphically depicted in Fig.~\ref{fig:nonstationary}(a), in which the
Q1 group is colored in blue and the Q4 group is colored in red.

In Fig.~\ref{fig:nonstationary}(b) and (c), we plot the MSDs for the
non-stationary ensembles Q1 (blue) and Q4 (red), along with the theoretical
MSD for the stationary state (black). The plots present the cases of (b)
the quenched extreme landscape and of (c) the annealed extreme landscape
($D_\mathrm{env}=0.5$), respectively.  At short times, the MSDs for the
Q1 and Q4 groups exhibit Fickian diffusion with the mean diffusivities
$\langle D_\mathrm{S}\rangle_\mathrm{Q1}$ and $\langle D_\mathrm{S}
\rangle_\mathrm{Q4}$, averaged over the Q1 and Q4 populations, respectively.
As a result, their MSDs are different from the theoretical MSD for the
stationary ensemble.

As time increases beyond the homogenization time, the tracer particles reach
the state of homogenization, and thus their MSDs (as well as the averaged
sampled diffusivity) converge to $2\langle D_\mathrm{S}\rangle t$, independent
of the particle initial positions. At the cross-over timescale, a transient
anomalous diffusion regime $\mathrm{MSD}\sim t^{\alpha}$ ($\alpha\neq1$)
emerges, bridging between the short and long-time Fickian diffusion
regimes. We also note that as the particles reach homogenization faster with
an increased value for $D_\mathrm{env}$, i.e., the long-time linear scaling
$\sim 2\langle D_\mathrm{S} \rangle t$ is observed earlier for
$D_\mathrm{env}=0.5$ than for $D_\mathrm{env}=0$.

\section{Calculation of the superstatistical integral}
\label{sec:appendixC}

\subsection{Exact solution of the superstatistical integral and asymptotic limit}

Given a diffusion constant $D_\mathrm{env}$, Eq.~\eqref{eq:superstatistics} is
expressed as follows (the normalization is omitted for convenience):
\begin{equation}
\label{eq:integral_appendix}
p(x_{\Delta},\Delta)\propto\int_0^{\infty}\mathrm{d}D\,\frac{\exp\left(-\frac{
x_{\mathit\Delta}^2}{4D\mathit\Delta}-\frac{D}{D_0}\right)}{\sqrt{4\pi D\mathit
\Delta}}\frac{1}{D+D_\mathrm{env}}.
\end{equation}
To evaluate this integral, we utilize the identity
\begin{equation}
\frac{1}{D+D_\mathrm{env}}=\int_0^{\infty}\mathrm{d}u\,\mathrm{e}^{-u(D+
D_\mathrm{env})}.
\end{equation}

\begin{figure}
\centering
\includegraphics[width=0.49\textwidth]{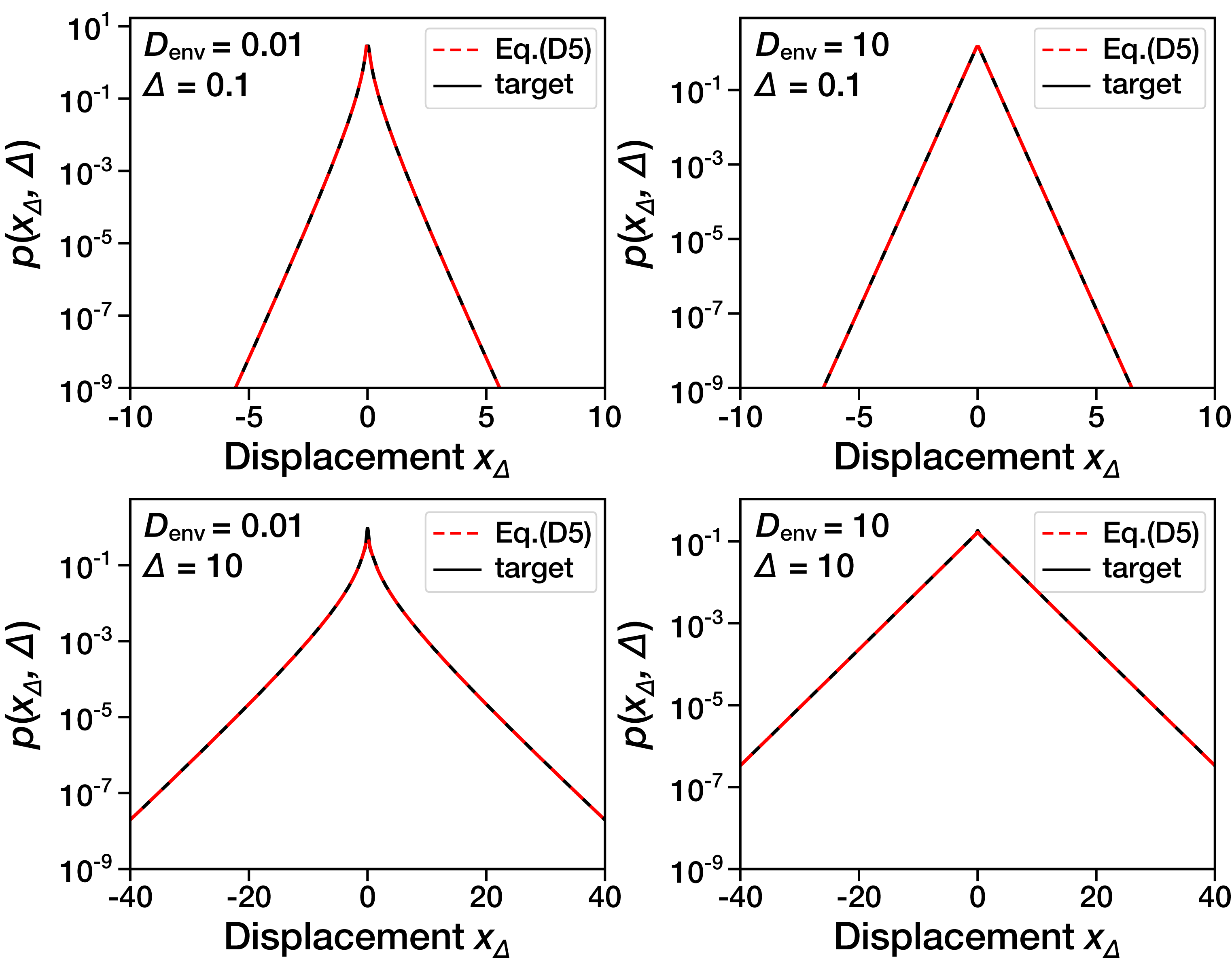} 
\caption{\textbf{Displacement PDFs for superstatistical integration,
Eq.~\eqref{eq:integral_appendix}.} Black solid lines represent the numerical
evaluation of Eq.~\eqref{eq:integral_appendix}, and red dashed lines show Eqs.~\eqref{eq:exact_appendix}. The red dashed lines perfectly match the black solid lines.}
\label{figa1}
\end{figure}

As both integrals are finite, we can safely switch the position of the two
integrals, which yields
\begin{eqnarray}
\nonumber
p(x_{\mathit\Delta},\mathit\Delta)&\propto&\int_0^{\infty}\mathrm{d}u\,\int_0
^{\infty}\mathrm{d}D\,\frac{ \mathrm{e}^{-\frac{x_{\mathit\Delta}^2}{4D \mathit
\Delta}-\frac{D}{D_0}}}{\sqrt{4\pi D\mathit\Delta}}\mathrm{e}^{-u(D+ D_\mathrm{
env})}\\
\nonumber
&\propto&\int_0^{\infty}\mathrm{d}u\,\mathrm{e}^{-uD_\mathrm{env}}\int_0^{
\infty}\mathrm{d}D\,\frac{\mathrm{e}^{-(u+\frac{1}{D_0})D}}{\sqrt{4\pi D\mathit
\Delta}}\\
&&\times\exp\left(-\frac{x_{\mathit\Delta}^2}{4D\mathit\Delta}\right).
\end{eqnarray}
From a change of variable $y=\sqrt{D}$, we obtain
\begin{eqnarray}
\nonumber
p(x_{\mathit\Delta},\mathit\Delta)&\propto&\sqrt{\frac{1}{\pi\mathit\Delta}}
\int_0^{\infty}\mathrm{d}u\,\mathrm{e}^{-u D_\mathrm{env}}\\
\nonumber
&&\times\int_0^{\infty}\mathrm{d}y\,\exp\left(-\left(u+\frac{1}{D_0}\right)
y^2\right)\\
\nonumber
&&\times\exp\left(-\frac{x_{\mathit\Delta}^2}{4\mathit\Delta y^2}\right)\\
\nonumber
&=&\sqrt{\frac{1}{4\mathit\Delta}}\int_0^{\infty}\mathrm{d}u\,\mathrm{e}^{-u
D_\mathrm{env}}\frac{1}{\sqrt{u+\frac{1}{D_0}}}\\
&&\times\exp\left(-2\sqrt{\left(u+ \frac{1}{D_0}\right)\frac{x_{\mathit\Delta}
^2}{4\mathit\Delta}}\right).
\end{eqnarray}
The change of variables $z=\sqrt{u+1/D_0}$ leads to the Gaussian integral
\begin{eqnarray}
\nonumber
p(x_{\mathit\Delta},\mathit\Delta)&\propto&\sqrt{\frac{1}{\mathit\Delta}}
\int_{D_0^{-1/2}}^{\infty}\mathrm{d}z\,\mathrm{e}^{-(z^2-\frac{1}{D_0})
D_\mathrm{env}}\mathrm{e}^{-z\sqrt{x_{\mathit\Delta}^2/\mathit\Delta}}\\
\nonumber
&=&\sqrt{\frac{\pi}{4D_\mathrm{env}\mathit\Delta}}\mathrm{erfc}\left(\frac{
|x_{\mathit\Delta}|}{2\sqrt{D_\mathrm{env}\mathit\Delta}}+\sqrt{\frac{
D_\mathrm{env}}{D_0}}\right)\\
&&\times\exp\left(\frac{x_{\mathit\Delta}^2}{4D_\mathrm{env}\mathit\Delta}+
\frac{D_\mathrm{env}}{D_0}\right).
\label{eq:exact_appendix}
\end{eqnarray}
This is the exact solution for our superstatistical integral presented in
Eq.~\eqref{eq:integral_appendix}. We now explore the asymptotic behavior of
the PDF $p(x_{\mathit\Delta},\mathit\Delta)$. In the asymptotic limit $
|x_{\mathit\Delta}|/[2\sqrt{D_{\mathrm{env}}\mathit\Delta}]+\sqrt{D_{
\mathrm{env}}/D_0}\gg1$, indicative for large $x_{\mathit\Delta}$, large
$D_{\mathrm{env}}$, small $D_{\mathrm{env}}$, or small $\mathit\Delta$, we
employ the approximation $\mathrm{erfc}(y)\approx\exp(-y^2)/[y\sqrt{\pi}]$
for large $y$. This yields
\begin{eqnarray}
\nonumber
p(x_{\mathit\Delta},\mathit\Delta)&\propto&\sqrt{\frac{1}{4D_\mathrm{env}
\mathit\Delta}}\exp\left(\frac{x_{\mathit\Delta}^2}{4D_\mathrm{env}\mathit
\Delta}+\frac{D_\mathrm{env}}{D_0}\right)\\
\nonumber
&&\times\frac{\exp\left(-\left[\frac{|x_{\mathit\Delta}|}{2\sqrt{D_\mathrm{
env}\mathit\Delta}}+\sqrt{\frac{D_\mathrm{env}}{D_0}}\right)^2\right)}{\frac{
|x_{\mathit\Delta}|}{2\sqrt{D_\mathrm{env}\mathit\Delta}}+\sqrt{\frac{
D_\mathrm{env}}{D_0}}}\\
&\propto&\frac{\exp\left(-\frac{|x_{\mathit\Delta}|}{\sqrt{D_0\mathit\Delta}}
\right)}{|x_{\mathit\Delta}|+2\sqrt{D_\mathrm{env}\mathit\Delta}\sqrt{\frac{
D_\mathrm{env}}{D_0}}}.
\label{eq:asymptotic_appendix}
\end{eqnarray}
We display this analytical solution in Fig.~\ref{figa1} and compare it with
the target PDF from Eq.~\eqref{eq:integral_appendix}. The target PDFs, shown
as black solid lines, were derived by numerically superimposing Gaussian PDFs weighted by $\psi(D)$. The perfect alignment of these plots across various
$D_{\mathrm{env}}$ and $\mathit\Delta$ values verifies our analytical solution.

\subsection{Approximation of a superstatistical integral with Gaussian PDFs}

\begin{figure}
\includegraphics[width=0.38\textwidth]{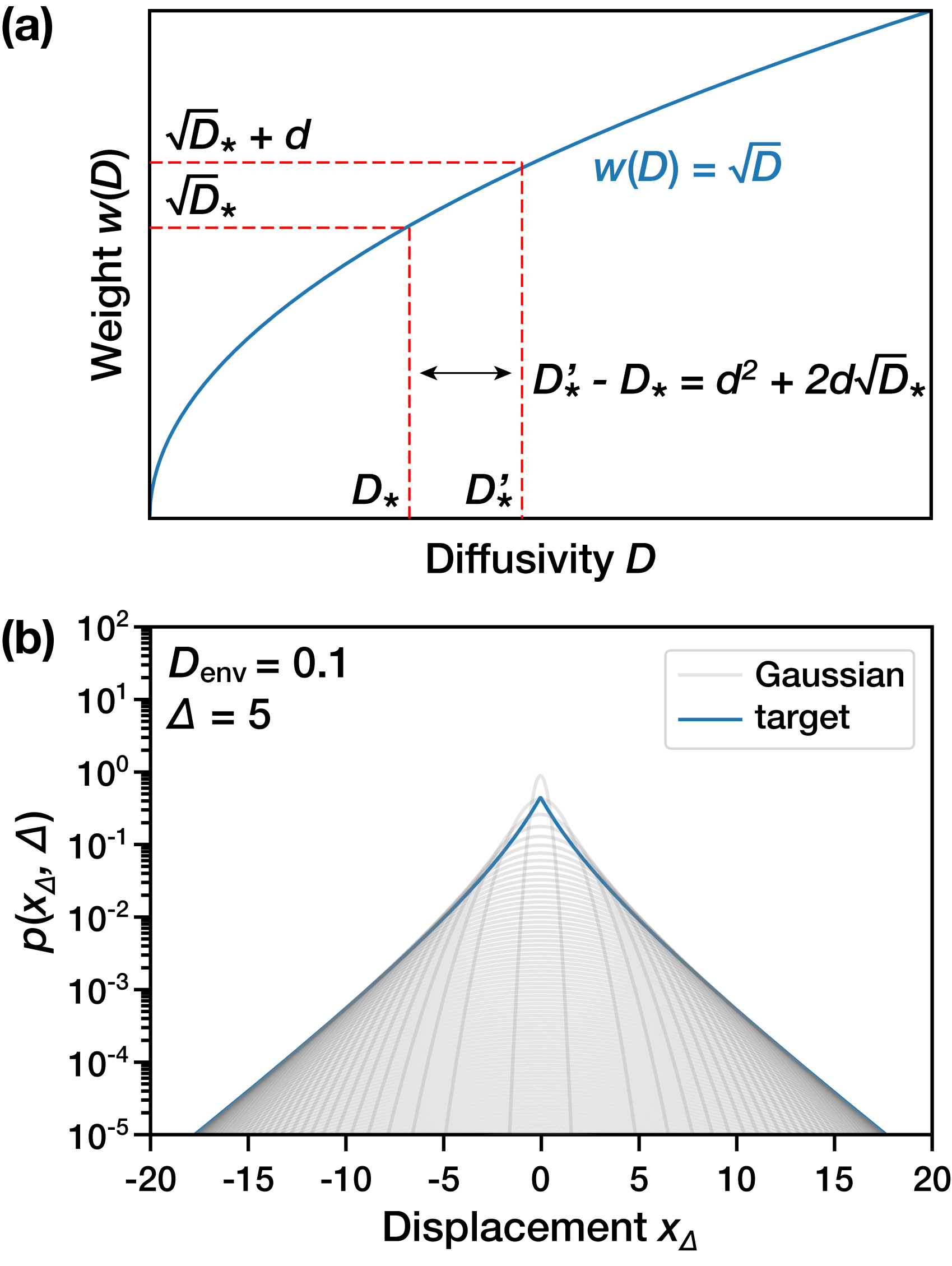} 
\caption{\textbf{Approximation of a superstatistical integral with Gaussian
PDFs.} (a) Characteristic width of Gaussian PDF vs. diffusivity $D$. (b)
Schematic illustration of the approximation for the displacement PDF $p(x_{
\mathit\Delta},\mathit\Delta)$. Gray solid lines are Gaussian PDFs with
diffusivity $D$, multiplied by the effective weight $\psi(D)\sqrt{D}$. Blue
solid line: numerically evaluated superstatistical integral
\eqref{eq:integral_appendix}.}
\label{figa2}
\end{figure}

Although we obtained the exact solution for $p(x_{\mathit\Delta},\Delta)$,
it is often nontrivial to calculate such kinds of superstatistical integrals.
In this Subsection, we introduce a method for approximating superstatistical
integrals. The basic idea of our approach is quite similar to the saddle
point approximation \cite{luo2019}. We consider the following integral
\begin{equation}
\label{eq:integral_appendix2}
p(x_{\mathit\Delta},\mathit\Delta)=\int_0^{\infty}\mathrm{d}D\,\psi(D)G(x_{
\mathit\Delta},\mathit\Delta|D).
\end{equation}
This integral implies that we superpose Gaussian PDFs $G(x_{\mathit\Delta},
\mathit\Delta|D)$ for a specific $D$, weighted by $\psi(D)$, as $D$ varies
linearly from 0 to infinity. We define the characteristic width of a
Gaussian PDF $(4\pi D\mathit\Delta)^{-1/2}\exp\left(-x_{\mathit\Delta}^2/[4D
\mathit\Delta]\right)$ as $w(D)=\sqrt{D}$. For a given $D=D_*$, consider its
characteristic width $w(D_*)=\sqrt{D_*}$ and some infinitesimally small
interval $d$ just above $\sqrt{D_*}$, as illustrated in Fig.~\ref{figa2}(a).
The length of the diffusivity domain that corresponds to characteristic
widths within the range $w(D)\in[\sqrt{D_*},\sqrt{D_*}+d]$ is given by $D_*'
-D_*=d^2+2d\sqrt{D_*}$. Consequently, for Gaussian PDFs with widths $w(D)$
in the range $[\sqrt{D},\sqrt{D}+d]$, the effective weight can be
approximated as
\begin{equation}
\psi(D)(d^2+2d\sqrt{D})\approx\psi(D)(2d\sqrt{D})\propto\psi(D)\sqrt{D}.
\end{equation}

We now define $D_\mathrm{max}(x)$ as the diffusivity that maximizes the
contribution at $x_{\mathit\Delta}=x$,
\begin{equation}
D_\mathrm{max}(x)=\mathrm{argmax}_D\left\{\psi(D)G(x_{\mathit\Delta}=x,
\mathit\Delta|D)\sqrt{D}\right\}.
\end{equation}
Since the value of a Gaussian PDF $G(x_{\mathit\Delta},\mathit\Delta|D)$ at
a fixed $x_\mathit\Delta$ varies significantly with changes in its parameter
$D$, the contribution from a Gaussian PDF with $D_\mathrm{max}(x_{\mathit
\Delta})$, represented as $\psi(D_\mathrm{max}(x_{\mathit\Delta}))G(x_{\mathit
\Delta},\mathit\Delta|D_\mathrm{max}(x_{\mathit\Delta}))\sqrt{D_\mathrm{
max}(x_{\mathit\Delta})}$, dominates the value of $p(x_{\mathit\Delta},
\mathit\Delta)$. Consequently, we can approximate $p(x_{\mathit \Delta},\mathit
\Delta)$ as
\begin{eqnarray}
\nonumber
p(x_{\mathit\Delta}=x,\mathit\Delta)&\approx&\mathscr{N}\psi[D_\mathrm{max}
(x)]G[x,\mathit\Delta|D_\mathrm{max}(x)]\\
&&\times\sqrt{D_\mathrm{max}(x)},
\label{eq:b10}
\end{eqnarray}
where $\mathscr{N}$ is a normalization factor.

We graphically illustrate the approximation process described above in
Fig.~\ref{figa2}(b). The target displacement PDF \eqref{eq:superstatistics}
for $D_{\mathrm{env}}=0.1$ is shown as the blue solid line. This PDF is
juxtaposed with the multiple Gaussian PDFs $\psi(D)G(x_{\Delta},\Delta|D)
\sqrt{D}$, each corresponding to different $D$ values, depicted as grey
solid lines. Essentially, the approximation in Eq.~\eqref{eq:b10} involves
deriving the outer envelope of these multiple Gaussians.

\begin{figure}
\centering
\includegraphics[width=0.49\textwidth]{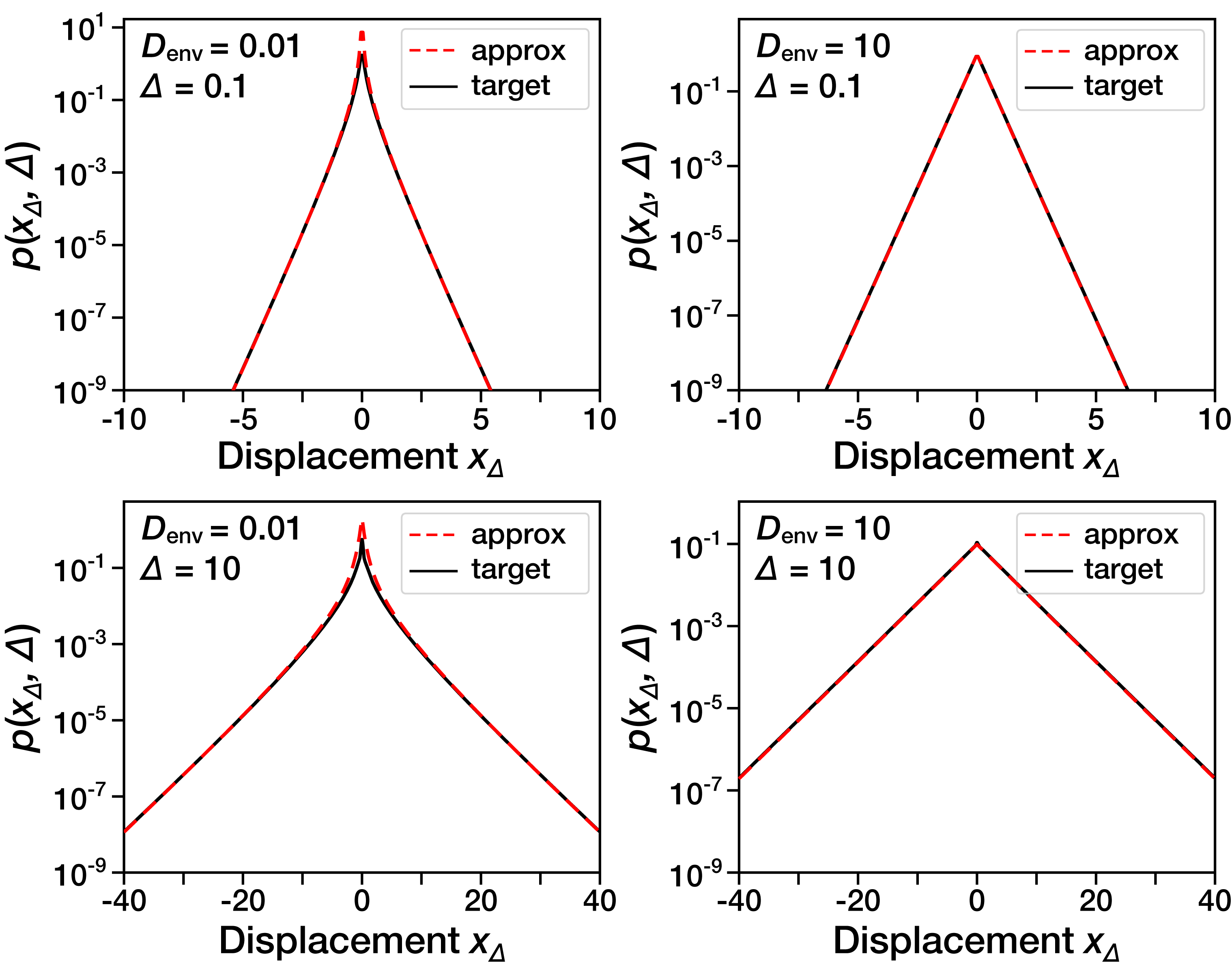} 
\caption{\textbf{Displacement PDFs for superstatistical integral
\eqref{eq:integral_appendix} with approximation \eqref{eq:approx_appendix}.}
Black solid lines are obtained from the numerical evaluation of
Eq.~\eqref{eq:integral_appendix}, and red dashed lines are plotted using
Eq.~\eqref{eq:approx_appendix}. The red dashed lines show a slight deviation
from the target PDF for small $D_\mathrm{env}(=0.01)$, while they perfectly
match the target PDF for large $D_\mathrm{env} (=10)$.}
\label{figa3}
\end{figure}

By inserting the weight function $\psi(D)\propto\exp(-D/D_0)/(D+D_\mathrm{
env})$, we find
\begin{equation}
\label{eq:approx_appendix}
p(x_{\mathit\Delta},\mathit\Delta)\approx\frac{\mathscr{N}'}{\sqrt{\mathit
\Delta}}\frac{\exp\left[-\frac{x_{\mathit \Delta}^2}{4D_\mathrm{max}(x_{
\mathit\Delta})\mathit\Delta}-\frac{D_\mathrm{max}(x_{\mathit\Delta})}{D_0}
\right]}{D_\mathrm{max}(x_{\mathit\Delta})+D_\mathrm{env}},
\end{equation}
where $\mathscr{N}'$ is a normalization factor and
\begin{eqnarray}
\nonumber
D_\mathrm{max}(x_{\mathit\Delta})&=&-\frac{1}{3}(D_\mathrm{env}+D_0)+\frac{
\left[\beta+(4\alpha^3+\beta^2)^{1/2}\right]^{1/3}}{12\cdot2^{1/3}}\\
&&-\frac{\alpha}{6\cdot2^{2/3}\left[\beta+(4\alpha^3+\beta^2)^{1/2}\right]^{
1/3}}
\end{eqnarray}
with
\begin{eqnarray}
\nonumber
\alpha&=&-16(D_\mathrm{env}+D_0)^2-\frac{12}{\mathit\Delta}D_0 x_{\mathit
\Delta}^2\\
\nonumber
\beta&=&-128D_\mathrm{env}^3-384D_0D_\mathrm{env}^2-384D_0^2D_\mathrm{env}
-128D_0^3\\
&&+\frac{288}{\mathit\Delta}D_0D_\mathrm{env}x_{\mathit\Delta}^2-\frac{144}{
\mathit\Delta}D_0^2x_{\mathit\Delta}^2.
\end{eqnarray}
From Eq.~\eqref{eq:approx_appendix}, one can derive the asymptotic forms
\begin{equation}
p(x_\Delta,\Delta)\approx \left \{ \begin{array}{ll}\frac{2\mathscr{N}'}{\sqrt{D_0}
|x_\Delta|}\exp\left(-\frac{|x_\Delta|}{\sqrt{D_0\Delta}}\right),&\frac{x_
\Delta^2}{\Delta}\gg D_0,D_\mathrm{env}\\ [1.2em]
\frac{\mathscr{N}'}{D_\mathrm{env} \sqrt{\Delta}}\exp\left(-\frac{|x_\Delta|}{
\sqrt{D_0\Delta}}\right),&\text{large}~D_\mathrm{env},
\end{array} \right.
\end{equation}
which are the same as Eq.~\eqref{eq:asymptotic_quenched} and \eqref{eq:asymptotic_laplace}.

We compare our result \eqref{eq:approx_appendix} with the target PDF in
Fig.~\ref{figa3}. For small $D_\mathrm{env}$ (=0.01, left panels), the tails
at $x_\Delta^2/\Delta\gg D_\mathrm{env}$) perfectly overlap with the target
PDF (black). For large $D_\mathrm{env}~(=10)$, the tails are well matched for
all cases, while the peaks deviate somewhat from the target PDF for small
$D_\mathrm{env}$.

\section{Non-Gaussianity parameter for finite-length trajectories}
\label{sec:appendixD}

The non-Gaussianity parameter (NGP) for a one-dimensional (1D), centered
random variable $X$ is typically defined as
\begin{eqnarray}
\nonumber
\mathrm{NGP}(X)&=&\frac{1}{3}\times\mathrm{Kurt}[X]-1\\
&=&\frac{1}{3}\times\frac{\mathbb{E}\left[X ^4\right]}{\mathbb{E}\left[X^2
\right]^2}-1.
\label{eq:conventionalNGP}
\end{eqnarray}
For Gaussian-distributed independent and identically distributed (i.i.d.)
random samples $X_i\sim\mathcal{N}(0,\sigma^2)$, the NGP is zero because the
kurtosis of a Gaussian PDF is three. However, when dealing with a finite
realization $\{X_i\}_{i=1}^N$ with $(N<\infty)$ of i.i.d.~random samples,
these samples do not follow an ideal Gaussian PDF due to the presence of a
cutoff. According to extreme value statistics \cite{gumbel1958statistics}, the
cutoff is $\mathbb{E}[\max X|N]=\sigma\sqrt{2\log N}$. The samples instead
follow a truncated normal distribution defined on $X\in[-\sigma\sqrt{2\log N},
\sigma\sqrt{2\log N}]$, resulting in a negative NGP.

For a finite number of random samples, we propose the modified NGP (mNGP) in
the form
\begin{equation}
\widehat{\mathrm{NGP}}(X;N)=\mathrm{NGP}(X)-\mathrm{NGP}_0.
\end{equation}
Here, $\mathrm{NGP}_0$ refers to NGP for the random variable $\xi_N^\mathrm{
tn}\in[-\sqrt{2\log N},\sqrt{2\log N]}$ of a truncated standard normal PDF,
where $N$ is the number of independent samples. If NGP for the random variable
$X$ is very similar to that of $\xi_N^\mathrm{tn}$, the mNGP of $X$ becomes
almost zero, and one can infer that the random samples $X$ are
Gaussian-distributed. Here, we used the truncated \textit{standard\/} normal
PDF because a truncated normal PDF with $\sigma$, $N$, and the domain $[-\sigma
\sqrt{2\log N},\sigma\sqrt{2\log N}]$ exhibits the same NGP values independent
of $\sigma$, solely depending on $N$. In our work, the second and fourth moments
of the truncated normal PDF, which are necessary to calculate $\mathrm{NGP}$,
are computed using the SciPy module in Python \cite{2020SciPy-NMeth}.

For a trajectory dataset with the observation time $T$ and particle number
$N_\mathrm{par}$, mNGP for the displacement of $x_\mathit\Delta$ is defined as
\begin{equation}
\widehat{\mathrm{NGP}}(x_\mathit\Delta;\mathit\Delta,T,N_\mathrm{par})=
\mathrm{NGP}(x_\mathit\Delta)-\mathrm{NGP}_0.
\end{equation}
Here, the $\mathrm{NGP}_0$ refers to the truncated normal random variables with
$N_\mathrm{par}T/\mathit\Delta$ independent displacement samples.

\begin{figure}
\includegraphics[width=0.38\textwidth]{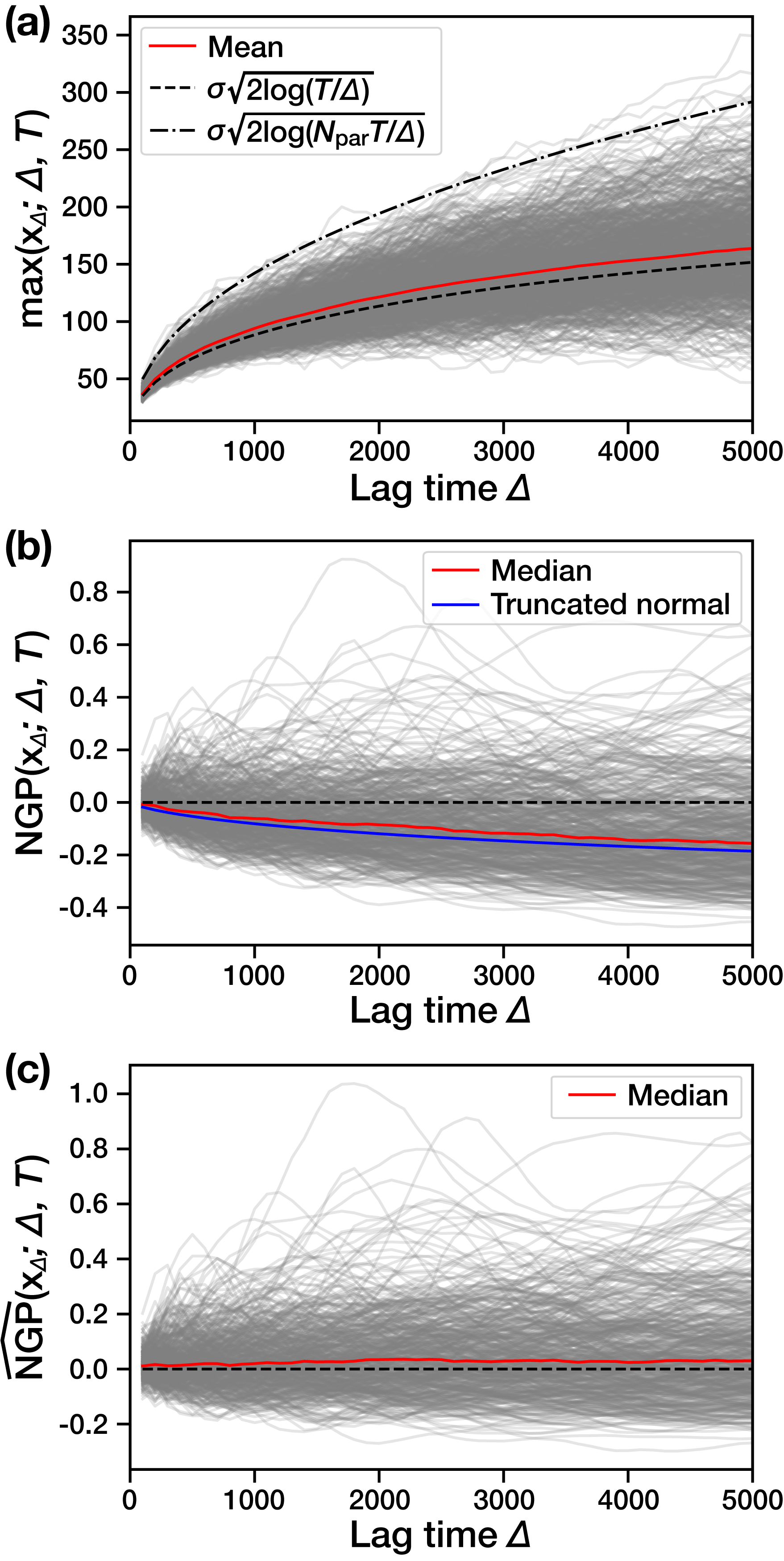} 
\caption{\textbf{Cutoff displacement and non-Gaussianity parameter of
finite-length Brownian trajectories for $D=0.5$, $N_\mathrm{par}=500$, and
$T=50000$).} (a) Maximal displacement as function of lag time ${\mathit\Delta}$.
The gray line represents the observed maximal displacement from individual
simulated trajectories, and their average is depicted by the red solid line.
Additionally, two theoretical values are depicted: (i) The upper dashed line is
the theoretical global maximum displacement for $N_\mathrm{par}$ trajectories.
(ii) The lower dashed line indicates the theoretical maximum displacement for a
single trajectory. (b) The conventional NGPs for individual trajectories (gray),
their median value (red), and the theoretical NGP for a truncated normal PDF,
$\mathrm{NGP}(\xi_{N_\mathrm{par}}^\mathrm{tn})$ (blue). The black dashed line
denotes the zero value. (c) mNGPs for individual trajectories (gray) and their
median value (red), the black dashed line shows the zero value.}
\label{figa4}
\end{figure}

In Fig.~\ref{figa4}, we test the validity of mNGP with stochastic simulations.
We simulated Brownian trajectories for which the unit-time displacement $x(t+1)
-x(t)$ was updated according to a Gaussian PDF $\mathcal{N}(0,2D)$. The
simulations were performed with $D=0.5$, $T=50000$, and $N_\mathrm{par}=500$.
For a given lag time $\mathit\Delta$, the variance of $x_\mathit\Delta$
satisfies $\langle x_\mathit\Delta^2\rangle=\mathit\Delta$.

First, Fig.~\ref{figa4}(a) presents the maximal displacement among $\{x(t+{
\mathit\Delta})-x(t)\}$ from a single trajectory for a given ${\mathit\Delta}$.
The gray lines are results from individual trajectories and the red solid line
shows their average value for the given ${\mathit\Delta}$. It is confirmed that
the averaged maximal displacement (red) is in excellent agreement with the
theoretical behavior $\sim\sigma\sqrt{2\log(T/\mathit\Delta)}$ (lower dashed
line) expected from extreme value statistics. We also observe that the
theoretical maximal displacement for $N_\mathrm{par}$ trajectories (upper
dashed line) explains very well the observed global maximal displacement
across all lag times investigated.

Second, we plot in Fig.~\ref{figa4}(b) the conventional NGP
\eqref{eq:conventionalNGP} for individual trajectories (gray) as a function
of $\mathit\Delta$. Their median values are depicted as the red solid line,
which agrees nicely with the theoretical curve for the NGP of a truncated
normal PDF with $T=50000$ and $N_\mathrm{par}=1$ (blue). Since NGP is
calculated from individual trajectories in the plot, the theoretical curve is
plotted with $N_\mathrm{par}=1$. Note that although we simulated a Gaussian
process, the finite Brownian trajectories result in a negative NGP, as
expected from our discussion.

Third, in Fig.~\ref{figa4}(c), we show mNGP for the simulated trajectories.
mNGP for individual trajectories is shown in gray, while their median value
is depicted in red. We demonstrate that, in contrast to the conventional NGP,
mNGP fluctuates around zero, indicating that it provides a more reliable
measure for Gaussianity as compared to the conventional NGP.

\section{Time-averaged mean squared displacement}
\label{sec:appendixE}

\begin{figure}
\includegraphics[width=0.49\textwidth]{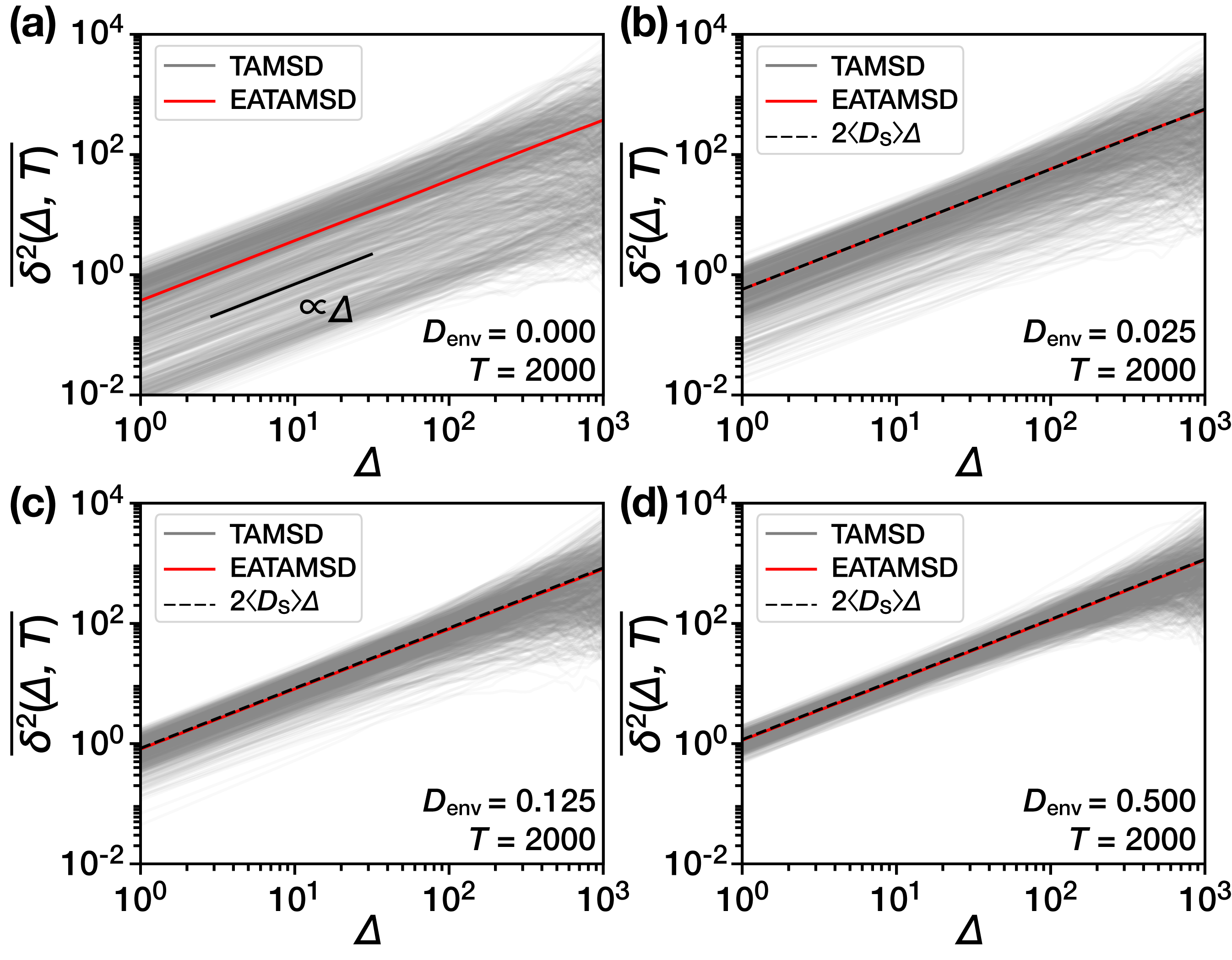} 
\caption{\textbf{TAMSD for short observation time $T=2\times10^3$.} In all
panels, TAMSDs from individual tracers (gray lines) are plotted together with
their average, i.e., the EATAMSD (black thick line). Eq.~\eqref{eq:fickianlaw}
is represented by the dashed line.}
\label{figa6}
\end{figure}

In Fig.~\ref{figa6} we show the TAMSD for short observation time, $T=2000$
(compare with Fig.~\ref{fig8}), much shorter than the homogenization times
($10^4\ldots10^5$) of the extreme landscapes we investigated in the main text.
The scatter in the amplitudes of the TAMSD from their average is substantial
as compared to the results in Fig.~\ref{fig8} with $T=2\times10^6$. 

\begin{figure}
\centering
\includegraphics[width=0.35\textwidth]{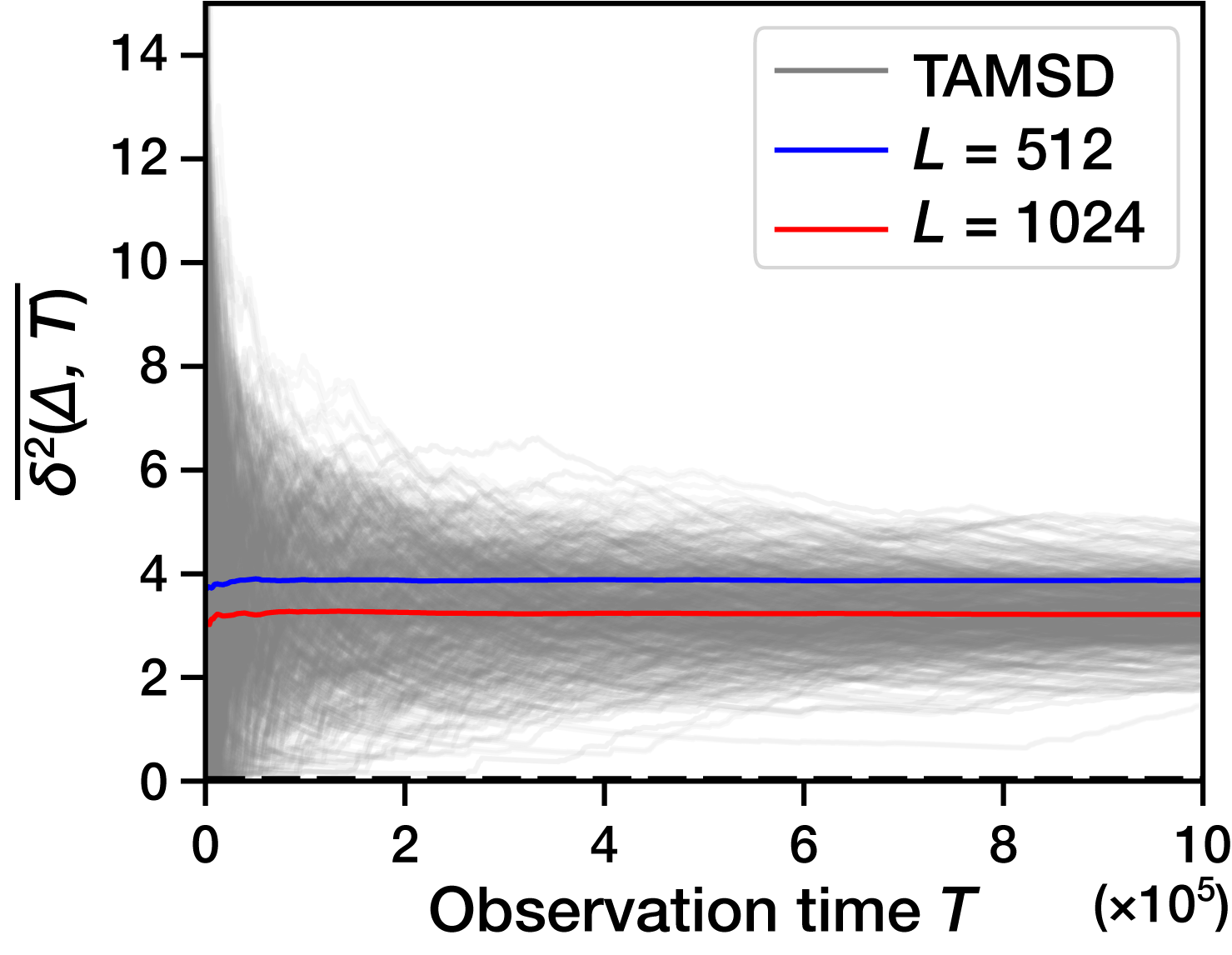} 
\caption{\textbf{TAMSD as function of observation time $T$ for $D_\mathrm{env}
=0$.} Gray solid lines are TAMSDs of tracers diffusing on a quenched environment
with $L=1024$, the red line represents their EATAMSD. For comparison, the EATAMSD
for $L=512$ is also shown (blue line).}
\label{figa7}
\end{figure}

Figure~\ref{figa7} shows the TAMSD as a function of the observation time $T$
in the quenched case, $D_\mathrm{env}=0$, for $L=1024$. We compare the EATAMSD
for this case (red line) with the EATMSD for $L=512$ (blue). The value of the
EATAMSD becomes smaller when a particle diffuses in a larger ($L=1024$)
environment, as anticipated in the discussion in Sec.~\ref{sec:EB}. As the size
of the environment $L$ becomes larger, deeper traps $V(\mathbf{r})$ appear.

\bibliography{references}

\end{document}